\pdfoutput=1 
\documentclass{JINST}
\usepackage{amsmath}                            
\usepackage{amsbsy}
\usepackage{booktabs}

\newcommand{\ud}{\mathrm{d}}

\newcommand{\br}{\vec {r}}
\newcommand{\bv}{\vec{v}}
\newcommand{\bE}{\vec {E}}

\newcommand{\Gerda}{\textsc{Gerda}}
\newcommand{\nubb}{$0\nu\beta\beta$}
\newcommand{\Qbb}{$Q_{\beta \beta}$}
\newcommand{\MaGe}{\textsc{MaGe}}
\newcommand{\GELATIO}{\textsc{GELATIO}}
\newcommand{\GF}  {\textsc{Geant4}}
\newcommand{\MGS} {\textsc{MGS}}

\title{
Signal modeling of high--purity Ge detectors with a small read--out electrode and
application to neutrinoless double beta decay search in $^{76}\mbox{Ge}$ 
}
\author{
M.~Agostini~$^{abcd}$\thanks{Corresponding author.}~, ~
C.~A.~Ur~$^b$, ~
D.~Budj\'{a}\v{s}~$^c$, ~
E.~Bellotti~$^{e}$, ~
R.~Brugnera~$^{bd}$, ~
C.~M.~Cattadori~$^{e}$, ~
A.~di Vacri~$^f$, ~
A.~Garfagnini~$^{bd}$, ~
L.~Pandola~$^f$, ~
and S. Sch\"{o}nert $^{ac}$\\
\llap{$^a$}~Physikdepartment E15, Technischen Universit\"at M\"unchen, M\"unchen, Germany\\
\llap{$^b$}~INFN Padova, Padova, Italy\\
\llap{$^c$}~Max-Planck-Institut f\"{u}r Kernphysik, Heidelberg, Germany\\
\llap{$^d$}~Dipartimento di Fisica dell'Universit{\`a} di Padova, Padova, Italy\\
\llap{$^e$}~INFN Milano Bicocca, Milano, Italy\\
\llap{$^f$}~INFN Laboratori Nazionali del Gran Sasso, Assergi, Italy\\\\
E-mail: \email{matteo.agostini@ph.tum.de}}
\abstract{%
The \Gerda~experiment searches for the neutrinoless double beta decay of
$^{76}\mbox{Ge}$ using high--purity germanium detectors enriched in
$^{76}\mbox{Ge}$. The analysis of the signal time structure provides a powerful
tool to identify neutrinoless double beta decay events and to discriminate them
from gamma--ray induced backgrounds. Enhanced pulse shape discrimination
capabilities of \emph{Broad Energy Germanium} detectors with a small read--out
electrode have been recently reported. This paper describes the full simulation
of the response of such a detector, including the Monte Carlo modeling of
radiation interaction and subsequent signal shape calculation.  A pulse shape
discrimination method based on the ratio between the maximum current signal
amplitude and the event energy applied to the simulated data shows quantitative
agreement with the experimental data acquired with calibration sources.  The
simulation has been used to study the survival probabilities of the decays which
occur inside the detector volume and are difficult to assess experimentally.
Such internal decay events are produced by the cosmogenic radio--isotopes
$^{68}\mbox{Ge}$ and $^{60}\mbox{Co}$ and the neutrinoless double beta decay of
$^{76}\mbox{Ge}$.  Fixing the experimental acceptance of the double escape peak
of the $2.614~\mbox{MeV}$ photon to $90\%$, the estimated survival probabilities
at $Q_{\beta\beta} = 2.039~\mbox{MeV}$ are ($86\pm3$)\% for $^{76}\mbox{Ge}$
neutrinoless double beta decays, ($4.5\pm0.3$)\% for the $^{68}\mbox{Ge}$
daughter $^{68}\mbox{Ga}$, and ($0.9^{+0.4}_{-0.2}$)\% for $^{60}\mbox{Co}$
decays.
}
\keywords{Detector modeling and simulations, Particle identification methods,
Gamma detectors}
\received{xx xx xx}
\accepted{xx xx xx}
\published{xx xx xx}
\begin{document}
\section{Introduction}
The search for neutrinoless double beta decay (\nubb) is the only
practical method to probe whether the neutrino is a Majorana particle, i.e. 
its own anti--particle. \nubb~decay
violates lepton number conservation and would establish new physics
beyond the Standard Model of particle physics. Provided that the
exchange of light Majorana neutrinos are generating the leading term
in \nubb~decay, effective neutrino masses down to a few tens of meV will
be probed in the next generation of \nubb~decay experiments.

One promising isotope is $^{76}\mbox{Ge}$. High--Purity
Germanium (HPGe) detectors can be produced from germanium material
enriched in $^{76}\mbox{Ge}$, which serves simultaneously as a source and as
detector medium.  The advantages of the intrinsic radio--purity and
excellent spectroscopic performance ($2-3 ~\mbox{keV}$ full width at half maximum
resolution at $Q_{\beta\beta} = 2039~\mbox{keV}$) of HPGe detectors have been
recognized early~\cite{bib:fiorini}, and until today, the most stringent experimental
limits on \nubb~decay have been achieved with germanium crystals enriched
in $^{76}\mbox{Ge}$~\cite{bib:HM}. 
Also one group reported a claim of evidence~\cite{bib:KK}.

The research work presented here was carried out in the framework of
the GERmanium Detector Array (\Gerda) experiment~\cite{bib:gerdaProposal} which recently started
its commissioning phase at the \emph{Laboratori Nazionali del Gran Sasso}
(LNGS) of the INFN in Italy. 
An array of bare HPGe detectors enriched to $86\%$ in
$^{76}\mbox{Ge}$ will be immersed in a cryogenic liquid which serves
simultaneously as a coolant for the germanium detectors and as a
high--purity shield against external radiation. The experiment pursues
a staged implementation: the mentioned claim of evidence will be
scrutinized in its first phase with about $15~\mbox{kg}\cdot\mbox{years}$ exposure
and a background index at \Qbb~of
${<10^{-2}}~\mbox{counts}/(\mbox{keV}\cdot \mbox{kg}\cdot \mbox{year})$. The
second phase of the experiment will explore half--lives up to $2\cdot
10^{26}$~years, with $100~\mbox{kg}\cdot\mbox{years}$ of exposure and background index
of ${<10^{-3}}~\mbox{counts}/(\mbox{keV}\cdot\mbox{kg}\cdot\mbox{year})$. 
Contingent on the results of the first two phases, a
third phase is conceived to probe half--lives $>10^{27}$ years. The
corresponding effective neutrino mass of $\gtrsim10~\mbox{meV}$ is predicted
based on the results of neutrino oscillation experiments assuming an inverted mass
hierarchy~\cite{bib:neutrinoOscillation}. 
To explore this parameter regime, an exposure of
several $1000~\mbox{kg}\cdot\mbox{years}$ and backgrounds
${<10^{-4}}~\mbox{counts}/(\mbox{keV}\cdot\mbox{kg}\cdot\mbox{year})$ are
required. To reach such background levels, which are about three
orders of magnitude below the best current values, novel techniques
are required which exploit the decay characteristics and topology of
background events.

In recent works, we could show that p--type HPGe detectors produced by
Canberra Semiconductor~\cite{bib:canberra} with a small
read--out electrode, referred to as thick--window \emph{Broad Energy Germanium} (BEGe)
detectors, exhibit pulse shape discrimination performance~\cite{bib:dusan1, bib:dusan2}
superior to coaxial HPGe detectors.
This allows to efficiently distinguish between single--site events (\nubb--like) and multi--site
events (gamma--ray backgrounds). Investigations of a bare BEGe detector in liquid
argon showed  unaltered performance as compared to operation in a vacuum cryostat~\cite{bib:marik}. 
An additional benefit of the small size of the read--out electrode is its low capacitance 
which results in a lower noise and therefore a superior resolution and a lower threshold 
compared to other types of large--volume HPGe detectors. 
Following a full production chain
test~\cite{bib:depBEGe} of BEGe detectors with modified isotopic 
composition (depleted in $^{76}\mbox{Ge}$), the \Gerda~collaboration decided to
adopt thick--window BEGe detectors as their baseline design for the
second phase of \Gerda.  R\&D with BEGe detectors is also carried out by
the Majorana collaboration~\cite{bib:MJBEGe,bib:barbeau}.

The scope of this paper is the ab--initio
modeling of the signal pulse shapes of BEGe detectors created by
particle interactions and the comparison with experimental data. The
energy depositions and event topology were generated with the
\GF~simulation package~\cite{bib:geant4, bib:geant4_2} and the
consequent signal pulse shape generated by drifting the charge carrier
in the calculated electric field. The characteristic pulse shapes of
the simulated pulses are then analyzed. First we validate the
simulation of pulse shapes and discrimination methods with
experimental data using calibration sources located
externally to the crystal. Then, we calculate the expected pulse shape
discrimination (PSD) cut efficiencies for $^{60}\mbox{Co}$, $^{68}\mbox{Ga}$ and
\nubb~decays internal to the detector. We also study the pulse shapes
of energy depositions close to the small read--out electrode and in its vicinity.

\section{Overview of the simulation}
\label{sec:simulation}
The simulation presented here replicates the physical processes 
involved in the generation of the Ge detector signals with application to the
particular case of Broad Energy Germanium (BEGe) detectors.
The simulation can be divided into three logical blocks: 
1)~interaction of ionizing particles with the Ge crystal; 
2)~collection of the charge carriers produced at each interaction point at the detector electrodes and  
time evolution of the induced signals; 
3)~signal shaping by read--out electronic devices.

The first block consists of a Monte Carlo simulation, including all the physical
processes involved in the passage of gamma--rays or charged particles through
matter. It provides the interaction points and the corresponding energy losses
within the crystal. This part of the simulation is performed within the 
\MaGe~framework~\cite{bib:mage}, which is based on the \GF~simulation
package~\cite{bib:geant4, bib:geant4_2}. 

The second block describes the dynamics of the charge carriers generated inside
the detector and provides the signals induced on the electrodes by the charge
movement. It is calculated by using a modified version of the Multi Geometry
Simulation (\MGS) software \cite{bib:mgs}.

The last block of the simulation takes into account the effects of the read--out 
electronic devices and of the electronic noise. It generates signals that can be
directly compared to the measured ones. 

\begin{figure}[tbp]
   \scriptsize
   \begin{center}
      \includegraphics{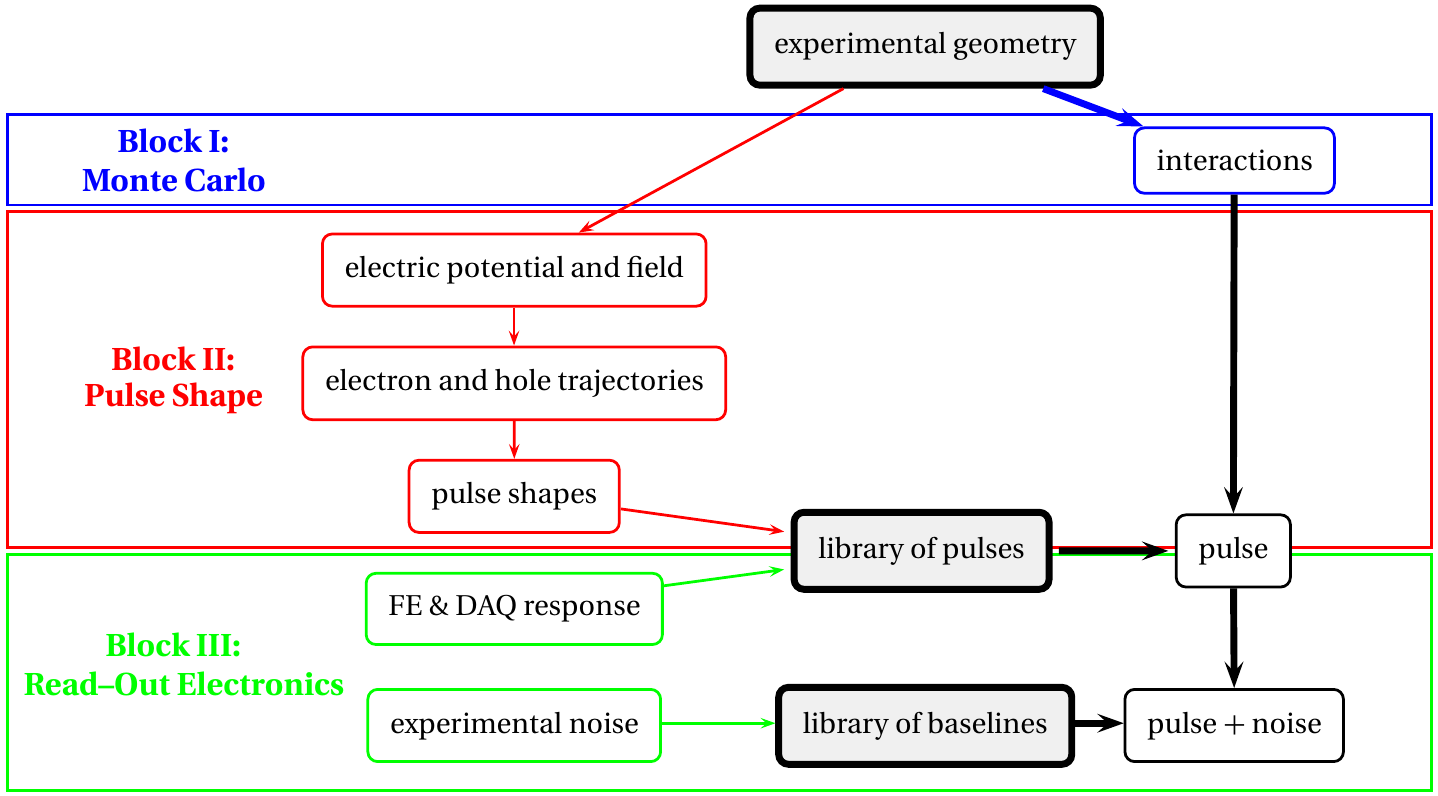}
   \end{center}
   \caption{Data flow of the simulation organized into three logical blocks.
   The data flow includes a first phase (thin arrows) in which  a library of
   pulses and baselines is created and a second phase (thick arrows) in which
   the libraries are used to generate the detector signals.
   In block III, FE \& DAQ refers to the preamplifier and the sampling device.
   }
   \label{fig:dataFlow}
\end{figure}

\figurename~\ref{fig:dataFlow} illustrates the implementation
diagram of these logical
blocks in the software. In the following two subsections the second and the
third block of the simulation are explained in more detail.

\subsection{Field calculation, charge carrier transport and signal generation}
\label{sec:PSS}
The pulse simulation performed with the \MGS-based software includes the
following steps:
1)~computation of the electric field inside the detector at defined potentials of the electrodes;
2)~transport of charge carries toward the electrodes;
3)~time evolution of the signal induced at the read--out electrode by the moving charges.

Since a semiconductor detector can be considered as an electrostatic system, 
the electric field can be computed by solving the Maxwell equations or,
equivalently, by solving the Poisson equation with a set of boundary
conditions for the potential.
In p--type Ge detectors, the p--n junction forms at the border of
the $\sim1~\mbox{mm}$ thick donor--doped surface layer (n+ layer). At operational
bias voltages the p--type volume is fully depleted of free charges (active
volume), while most of the n+ layer retains electrons in the conduction band
(anode, forming a dead layer). In our simulation we approximate this situation
by defining the active volume with a negative space charge distribution $\rho$
proportional to the net p--type impurity distribution, adjoined to the dead layer
on which the bias voltage is applied\footnote{%
Since our field calculation is performed on a $0.5~\mbox{mm}$ step grid, it can
not describe the thin part of the n+ volume in
which the conduction--band electrons are depleted (the n--side of the p--n
junction, with positive space charge).%
}. 
The boundary conditions are then provided by the value of the
potential on the conductive surfaces: $V_{cathode}$ and $V_{anode}$. 
Therefore the system of equations to solve is:
\begin{eqnarray}
   \left\{ 
   \begin{array}{lll}
      \nabla^2 \phi (\br) &=& - {\rho(\br)}/{(\varepsilon_{0} \varepsilon_{r}})\\
      \phi|_{S_{cathode}} &=& V_{cathode} \\
      \phi|_{S_{anode}}   &=& V_{anode}  \\
   \end{array}
   \right. 
\end{eqnarray}
where $\br$ is the position, $\phi (\br)$ is the electric potential, $\rho
(\br)$ is the charge density distribution, $\varepsilon_{0}$
is the vacuum electrical permittivity, $\varepsilon_{r}\sim16$ is the relative
permittivity of Ge, $\phi|_{S}$ is the potential at the surface $S$ surrounding
the considered electrode. 

The movement of charge carriers within the active volume is computed by using two phenomenological
models~\cite{bib:mihailescu, bib:bruynell01} which provide the drift velocity as a
function of the electric field magnitude and of the
electric field direction relative to the crystallographic axes.
We assume that each transfer of energy to the Ge crystal lattice results in the
generation of a cloud of free electrons and holes, which further drift as two
independent clusters. These are approximated in our simulation by two point-like
charges with opposite sign.
The trajectories are calculated with a fourth order Runge-Kutta method with
$1~\mbox{ns}$ time step.

Using the simulated trajectories of charge carriers and the weighting potential\footnote{%
The \emph{weighting potential} is a dimension--less quantity defined as the electric potential calculated
when the considered electrode is kept at a unit potential, all other electrodes
are grounded and all charges inside the device are removed.%
} distribution inside the detector, the charge signal $Q(\br(t))$ induced on the
electrodes is computed by using the Shockley--Ramo theorem~\cite{bib:shockleyramo}:
\begin{eqnarray}\label{eq:SR}
   Q(\br(t)) = -q_{tot} \, \phi_w (\br(t))
\end{eqnarray}
where $\br(t)$ is the position of the charge cluster at the time $t$, $q_{tot}$ the total
charge of the cluster and $\phi_w (\br(t))$ is the weighting potential.

The total signal of a simulated particle event composed of several interactions
is calculated as a sum of pulses from each hit, whose position and energy
deposition are provided beforehand with the \MaGe~Monte Carlo simulation. 
To reduce the computation time, the generation of the signals corresponding to
the individual hits is performed by using a library of pre--calculated pulses.
The library is generated by dividing the detector active
volume in $1~\mbox{mm}^3$ cubic cells, and simulating a single normalized
interaction at each corner.
The library must be
generated only once for each simulated detector geometry and bias voltage
setting. Then each Monte Carlo generated interaction is associated to one of the
cubic cells in the pulse library. The signal is computed as the weighted
average of the eight pulses in each corner of the cube, where the weight is
given by the inverse of the cubic euclidean distance between the interaction
point and the considered corner. The amplitudes of the individual interaction
pulses are then normalized according to the energy depositions in the hits and
all the pulses of the event are added up to one combined signal. 

The use of a pulse library decreases significantly the processing time of the
simulation in such a way that the pulse computation time is comparable with
that of the Monte Carlo simulation. 

\subsection{Read--out electronics response and noise}
\label{sec:DAQ}
The read--out electronics includes the preamplifier and the digital sampling
device. The response function was determined by providing the
preamplifier input with an impulse generated by a high--precision pulser, and then
by deconvolving the digitally sampled signal with the input signal. The simulated
detector signals are consequently convolved with this response function. This is
performed for each pulse in the library mentioned in previous subsection, so
that it needs to be done only once for each simulated experimental setup.

To reproduce the electronic noise present in the experimental data, samples are
taken from a library of experimentally recorded baselines. The amplitude of the
noise is normalized according to the experimental signal--to--noise ratio, and
the noise sample is then added to the calculated full event signal.
The output of the simulation is a file
of signals similar to those recorded experimentally with a digital data
acquisition system. Therefore it is possible to apply the same analysis tools to
both experimental and simulated data.

\section{Modeling of BEGe detectors}
\label{sec:BEGeModeling}
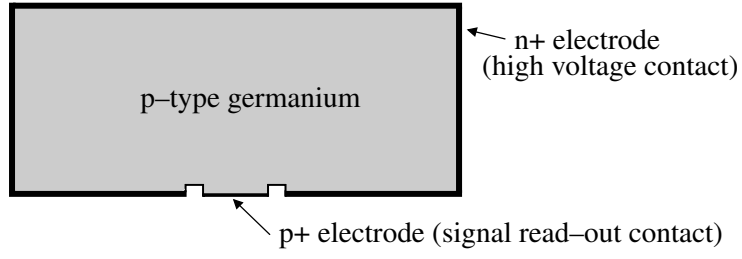
\begin{figure}[tb]
  \begin{center}
    {
      \includegraphics{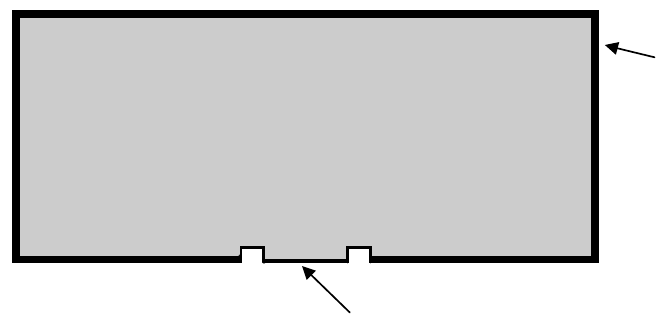}
      \put(-140,55){\makebox(0,0)[l]{\strut{p--type germanium}}}
      \put(-88,4){\makebox(0,0)[l]{\strut{p+ electrode (signal read--out contact)}}}
      \put(-0,77){\makebox(0,0)[l]{\strut{n+ electrode}}}
      \put(-12,67){\makebox(0,0)[l]{\strut{(high voltage contact)}}}
    }
  \end{center}
  \caption{Sketch of a BEGe detector. The signal read--out electrode and the
  groove are not to scale. The plot shows a vertical section of the
  detector passing through the symmetry axis. }
  \label{fig:BEGe}
\end{figure}

The detector used for the present simulation is a modified thick--window version of a \emph{Broad Energy
Germanium} (BEGe) detector~\cite{bib:BEGe}, a standard p--type
High--Purity Ge (HPGe) detector offered by Canberra
Semiconductor.  These detectors have a cylindrical shape
and a small B--implanted p+ electrode on one of the flat surfaces.  The
Li--diffused n+ electrode (between $0.4~\mbox{mm}$ and $0.8~\mbox{mm}$ in the
``thick window'' modification) covers the rest of the outer surface and is
separated from the p+ electrode by a circular groove.  The detector is mounted in
a $1.5~\mbox{mm}$ thick cylindrical aluminum housing. The positive high voltage
is applied to the n+ electrode while the signal is read out from the p+ electrode
where the holes are collected.  For the purpose of the present work the
specific case of the BE3830/S model was considered.  The BE3830/S model
has a diameter of $71~\mbox{mm}$ and a thickness of $32~\mbox{mm}$. The detector
was experimentally characterized in Refs.~\cite{bib:MatteoThesis,bib:assunta}.
A schematic drawing of a typical BEGe detector is shown in \figurename~\ref{fig:BEGe}.

In this section the signal formation and development in the BEGe detectors will
be discussed following the steps outlined in Section~\ref{sec:PSS}.  Moreover, the results
of the simulation will be used to show how it is possible to distinguish between
single--site and multiple--site events by analyzing the pulse shapes.

\subsection{The electric field inside BEGe detectors}
\label{sec:BEGeModelingField}
The pulse shape discrimination properties of BEGe detectors have
been attributed to the peculiar internal electric field which is created by the small
size of the read--out electrode. In Section~\ref{sec:PSS} the space charge distribution
inside the active volume (equivalent to the net distribution of acceptor
impurities) was identified along with the electrode potentials as a source of
electric field inside the detector.  To better understand the two contributions,
the linear superposition principle can be used to separate the potential into
the two individual components:
\begin{eqnarray}
    \phi(\br) = \phi_0(\br) + \phi_{\rho}(\br)
\end{eqnarray}
where $\phi_0$ is the potential calculated considering only the electrode
potentials and no impurity charge ($\rho (\br) =0 \,\,\,\, \forall \br$) and $\phi_{\rho}$ is the
potential generated by the impurity charge distribution when grounding all the
electrodes.
Therefore, we can solve the following two problems and then add up the solutions:
\begin{eqnarray}
\left\{
 \begin{array}{lll}
    \nabla^2 \phi_0(\br)  &=&  0 \\
    \phi_0|_{S_{cathode}} &=& V_{cathode} \\
    \phi_0|_{S_{anode}}   &=& V_{anode} \\
 \end{array}
\right. \qquad \qquad
\left\{
 \begin{array}{lll}
    \nabla^2 \phi_{\rho}(\br)  &=&  - \rho(\br) /( \varepsilon_{0} \varepsilon_{r}) \\
    \phi_{\rho}|_{S_{cathode}} &=&  0 \\
    \phi_{\rho}|_{S_{anode}}   &=&  0 \\
 \end{array}
\right.
\end{eqnarray}
Similarly, since the electric field is determined by the linear relation
$\bE = - \nabla \phi$, it can be also separated into two components:
\begin{eqnarray}
\bE(\br) = \bE_0(\br) + \bE_{\rho}(\br)
\end{eqnarray}
where $\bE_0(\br) = - \nabla \phi_0(\br)$ and
$\bE_{\rho}(\br) =  - \nabla \phi_{\rho}(\br)$.

\begin{figure}[t]
   \scriptsize
   \begin{center}
      \input{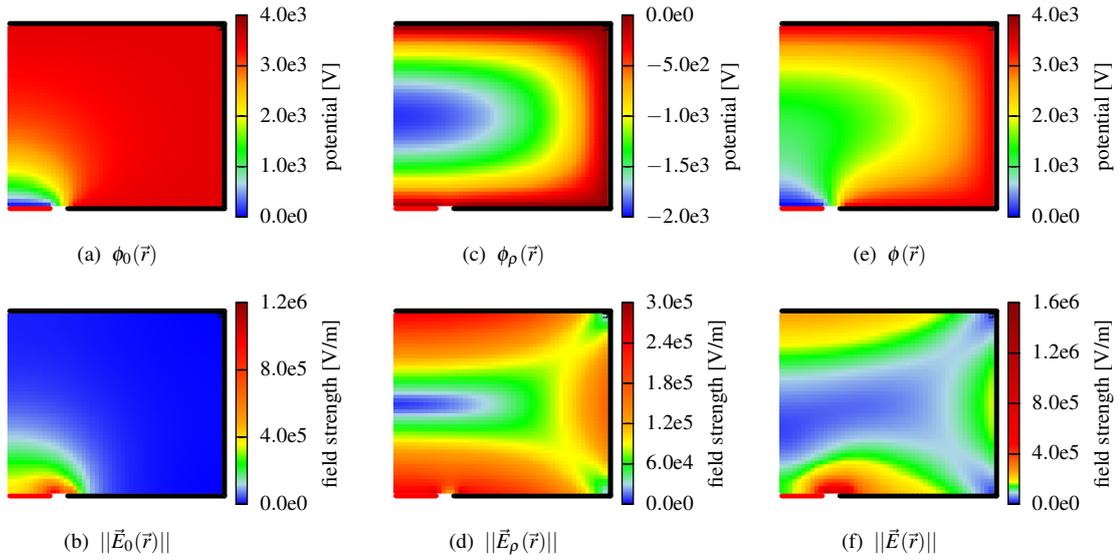}
   \end{center}
   \caption{Simulated electric potential and electric field strength for different
   configurations of a BEGe detector. In (a) and (b) the electrode
   potential is considered, in (c) and (d) the charge distribution, and in
   (e) and (f) the sum of the two contributions.
   The plots show half of a vertical section of the detector passing through the
   symmetry axis. The cathode is drawn in red and the anode in black.
   }
   \label{fig:efield}
\end{figure}

\figurename~\ref{fig:efield} shows the electric potential and the electric field
strength of the two
contributions and their sum for the BE3830/s detector operated in its
nominal configuration, i.e. cathode grounded, anode set at
$3500~\mbox{V}$ and $\rho~\sim~10^{10}~\mbox{impurity
atoms}/\mbox{cm}^3$.

\figurename~\ref{fig:efield}.a and ~\ref{fig:efield}.b show the electric
potential and field strength generated only by the electrodes ($\phi_0(\br)$ and
$||\bE_0(\br)||$).
The potential and the field show a sharp variation in the region close to the
small--size p+ electrode.
The field in the rest of the volume is so weak that the collection time would
be longer than the characteristic recombination time and most of the charge
would be lost.

\figurename~\ref{fig:efield}.c and ~\ref{fig:efield}.d show the electric
potential and field strength provided only by the impurity charge distribution
($\phi_{\rho}(\br)$ and $||\bE_{\rho}(\br)||$).  For the considered BEGe detector the charge
concentration is approximated by a uniform distribution.  The resulting negative
electric potential reaches its peak value in the middle of the detector. The
direction of the electric field in the region close to the small cathode is
opposite to that of $\bE_0$, whereas in the rest of the detector the field helps
to move the charges produced close to the outer n+ electrode towards the central
slice of the detector.

\figurename~\ref{fig:efield}.e and ~\ref{fig:efield}.f show the total electric
potential and field strength ($\phi(\br)$ and $||\bE(\br)||$).
The potential close to the small p+ electrode is dominated by the electrodes contribution
smoothed out by the opposite contribution of the impurity charge distribution field.
In the rest of the volume the dominant contribution is
provided by $E_{\rho}$. The effect of the $E_{\rho}$ field is to bring the holes in the
center of the detector while the field $E_0$ subsequently collects them to the
read--out electrode. This peculiar way of charge transportation in BEGe detectors
leads to the favorable signal shape as it 
will be discussed in the next sections.

\subsection{Signal development in dependence of the interaction position}
\label{sec:BEGeModelingWeightingField}
\begin{figure}[t]
  \begin{center}
      \scriptsize
      \input{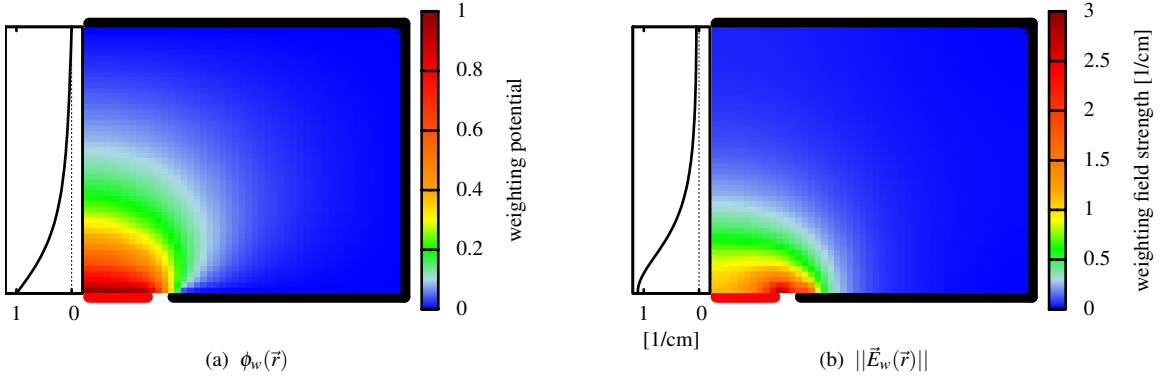}
  \end{center}
  \caption{%
     Weighting potential (a) and weighting field strength (b) of the small
     read--out electrode (cathode) computed for
     half of a vertical section of the BEGe detector passing through the detector symmetry
     axis (density maps) and for the symmetry axis (black lines in the side
     plots). 
     The cathode is drawn in red and the anode in black.
   }
   \label{fig:weighting}
\end{figure}

As given by equation~\eqref{eq:SR}, the weighting potential determines
the charge that a cluster of charge carriers drifting inside the detector 
induces on the read--out electrode.
\figurename~\ref{fig:weighting} shows the weighting potential $\phi_w(\br)$ and
the strength of the weighting field $||\bE_w(\br)|| = ||- \nabla \phi_w(\br)||$ of the
cathode in a BEGe detector.
The weighting potential (similarly to $\phi_0$ in \figurename~\ref{fig:efield}.a)
has a sharp variation in the region close to the small--size p+ electrode and it
is very weak in the rest of the detector volume (blue area in
\figurename~\ref{fig:weighting}).
If an interaction occurs in the volume of weak $\phi_w$, according to
equation~\eqref{eq:SR}, the signal induced by the holes drifting towards the small cathode
remains small until the charges arrive at about $1~\mbox{cm}$ away from the
electrode and then rapidly grows until the holes are collected.

On the other hand, in most of the active volume the electron contribution will be
present only at the beginning of the pulse, since electrons are collected to the
outer n+ electrode and the portion of the detector volume that is close to the outer
surface is much larger than the portion close to the p+ electrode. 
This effect is further
enhanced due to the fact that electrons move with 
velocities roughly two times higher than the holes. 
Moreover, the signal induced by the electron drifting will be
relatively small, as most electron clusters move in the region of weak
$\phi_w$. Thus, the contribution of the electrons to the formation of
the signals is expected to be negligible for most of the interaction positions.

\begin{figure}[t]
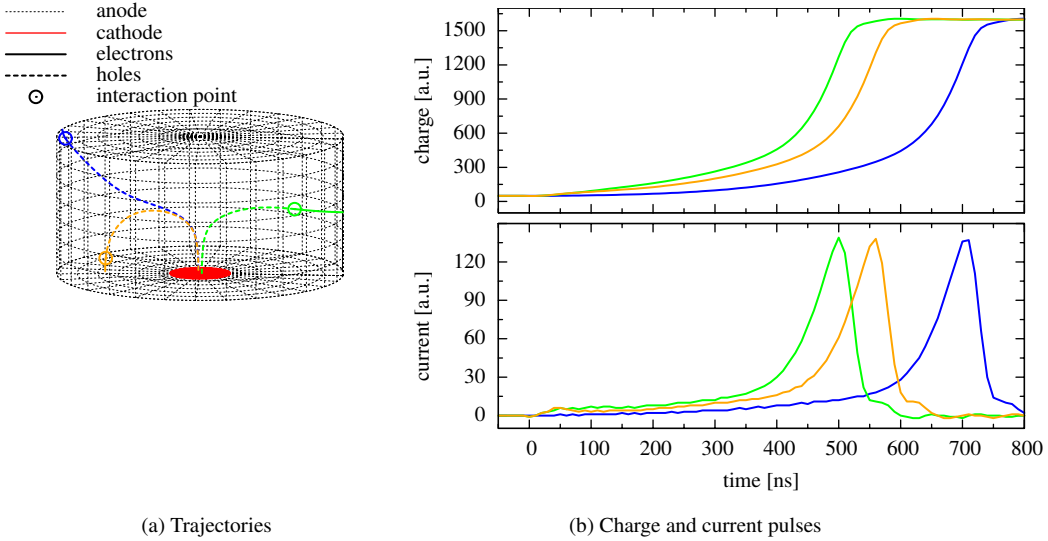

  \begin{center}
      \scriptsize
      \input{BEGeModelingWeightingTrajectories.tex}
      \put(-80,-15){\makebox(0,0)[r]{\strut{}(a) Trajectories}}%
      \input{BEGeModelingWeightingPulses.tex}
      \put(-100,-15){\makebox(0,0)[r]{\strut{} (b) Charge and current pulses}}%
  \end{center}
  \caption{%
     Simulated  electron--hole trajectories (a) and corresponding charge and current pulses (b)
     for three events occurring in different places in the bulk detector volume far from the p+
     electrode (``type I'' trajectories).
     The small oscillation in the current signals after the peak originates from the
     experimentally measured FE and DAQ response the pulses are convolved with.
     }
  \label{fig:weightingTrajectories}
\end{figure}

\figurename~\ref{fig:weightingTrajectories}.a shows the electron and hole trajectories
for three interactions in the ``bulk'' detector volume far from the p+
electrode. The holes follow the electric field (\figurename~\ref{fig:efield}.f)
and are first transported into the middle slice of the detector, then drift towards
the center of the detector and finally their trajectory bends towards the read--out electrode. 
It can be noticed that for all
these events, the last part of the hole collection happens along a common
path which is independent of the starting position. Accordingly, the last part
of the induced signal is identical for the different events. 
Since the first part of the signal, induced by the
holes and electrons in weak $\phi_w$ regions, is comparatively small, the signal
shapes are essentially independent of the interaction position. This can be
clearly seen in \figurename~\ref{fig:weightingTrajectories}.b, which shows the charge and
current pulses corresponding to the trajectories from
\figurename~\ref{fig:weightingTrajectories}.a.
The only visible difference is the
time shift of the rising part, which depends on the length of the charge carrier
path to reach the strong $\phi_w$ region. It is important 
to stress that this type of almost
indistinguishable signals originates from interactions in most 
of the detector volume, including corner regions. For further discussion we 
will refer to such signals as ``type I'' trajectories.

The shape of the current signals in \figurename~\ref{fig:weightingTrajectories}.b can be
understood by differentiating the function in equation~\eqref{eq:SR}. The current
induced at the cathode by a charge carrier is then given by:
\begin{eqnarray}\label{eq:SRC}
    I(\br(t)) =  \frac{\ud Q(\br(t))}{\ud t} 
    = q_{tot} \,\, \bv(\br(t)) \cdot \bE_w (\br(t))
\end{eqnarray}
From this equation we can see that the induced current $I(\br(t))$ depends on the velocity
$\bv (\br(t))$ and the weighting field $\bE_w (\br(t))$ at the position $\br(t)$
of the charge cluster. The
charge carrier velocity can vary roughly between $5~\mbox{cm/s}$ and
$10~\mbox{cm/s}$~\cite{bib:MatteoThesis}, while $E_w$ can increase by more than
a factor of 20 close to the read--out electrode (see
\figurename~\ref{fig:weighting}.b).
It is evident that $E_w$ has a dominant effect on the current signal -- the
signals of ``type I'' trajectories in
\figurename~\ref{fig:weightingTrajectories}.b feature
a significant current peak at the end of the hole collection.

\begin{figure}[t]
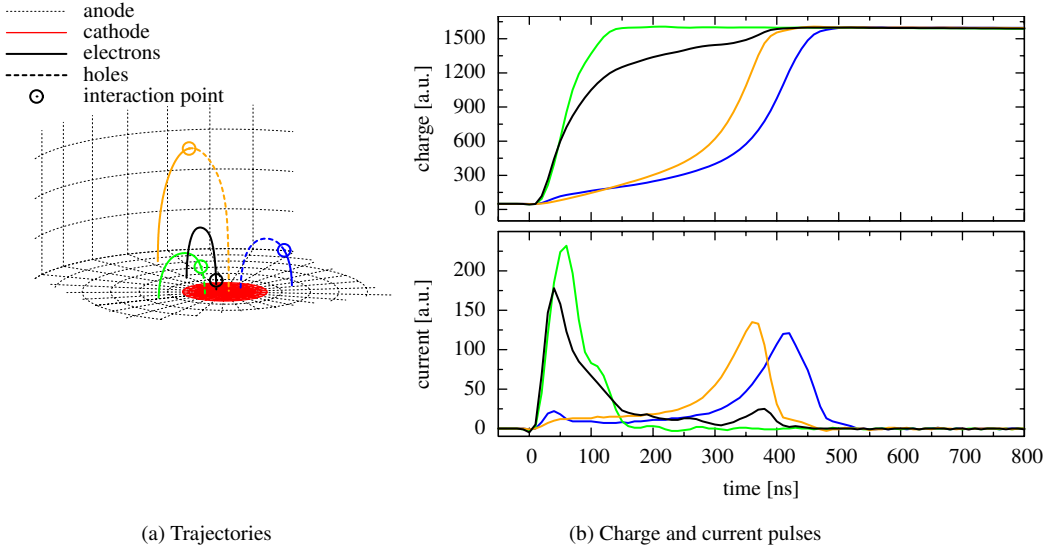

  \begin{center}
     \scriptsize
     \input{BEGeModelingAnomalousTrajectoriesZoom.tex}
      \put(-80,-15){\makebox(0,0)[r]{\strut{}(a) Trajectories}}%
     \input{BEGeModelingAnomalousPulses.tex}
      \put(-100,-15){\makebox(0,0)[r]{\strut{} (b) Charge and current pulses}}%
  \end{center}
   \caption{%
      Simulated electron--hole trajectories (a) and corresponding charge and current
      pulses (b) for three events occurring next to the cathode and a ``type I''
      event (orange line). For the green and black events (``type II''
      trajectories) both the electron and hole contribution are present at the
      very beginning of the signal. In the blue event (``type III'') the
      electron collection is quick and provides a characteristic kink in the
      first part of the signal.
   }
   \label{fig:weightingTrajectories2}
\end{figure}

Two other types of less common trajectories can be identified when the interactions occur
in the close vicinity of the groove and the small p+ electrode. Examples of
these kind of events are displayed in \figurename~\ref{fig:weightingTrajectories2}.
The ``type II'' trajectories originate close to the p+ electrode (green and black
color). In these events the holes are directly and quickly collected at the
cathode. Now also the electrons drifting in the opposite direction provide a
significant contribution to the signal, since they are now moving in a
region of strong $\phi_w$. The closer the interaction occurs to the cathode, the
more important the signal induced by electrons becomes. The signal is fully
dominated by the electron contribution for interactions within $\sim 2~\mbox{mm}$
from the p+ electrode (black example in
\figurename~\ref{fig:weightingTrajectories2}).
The induced charge signal rises quickly at
the beginning and then, as they drift away from the cathode into the weaker
$\phi_w$ regions, the signal growth slows down. The current peak appears at the
very beginning of the collection time. For events occurring few mm to
$\sim 1~\mbox{cm}$ from the p+ electrode (green example in
\figurename~\ref{fig:weightingTrajectories2}) neither electrons nor holes traverse the
full thickness of the strong $\phi_w$ region. So the main part of the signal is induced in a
relatively short time and the rise time is thus faster. The current peak is
amplified because contributions from both charge carrier types add up. The
current amplification can be further enhanced if the interaction happens close
to the inner edge of the groove, because here $E_w$ is strongest
(\figurename~\ref{fig:weighting}.b).

Starting points in a zone close to the anode, $\sim 1.5~\mbox{cm}$ from the
detector symmetry axis, result in ``type III'' trajectories (shown blue in
\figurename~\ref{fig:weightingTrajectories2}). For these events the electrons are
collected quickly, and since $\phi_w$ is still noticeable in this region, they
provide a characteristic kink in the first part of the signal. This quick
increase at the beginning causes the 10\% to 90\% rise time measurement
to give higher values for these events than for the ``standard'' type I
trajectory events.

\begin{figure}[t]
   \scriptsize
   \begin{center}
      \input{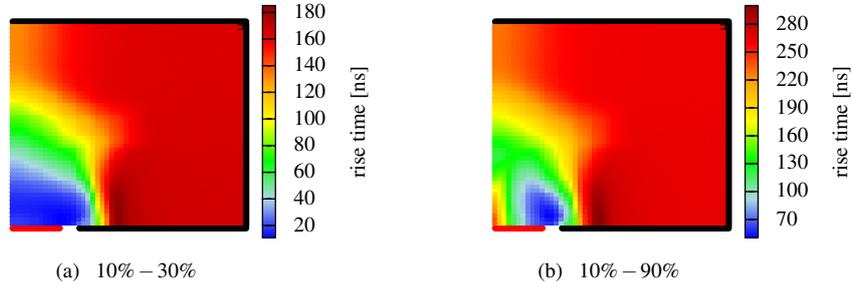}
   \end{center}
   \caption{Signal rise times 10--30\% (a) and
   10--90\% (b) as a function of the interaction position for a  BEGe detector.
   }
   \label{fig:risetime}
\end{figure}

The three types of signals and the extent of the volumes from which they
originate can be better understood from the plots shown in
\figurename~\ref{fig:risetime}.
The figure shows the rise times in the 10\% to 30\% and 10\%
to 90\% intervals of the signal height, as a function of the interaction position.
The type II trajectories originate in the volume close to the p+ electrode
($\sim$6\% of the active volume),
distinguished by short rise times plus a small zone of longer rise times few mm
away from the center of the electrode. 
The type III trajectory starting points
can be identified in the area of increased rise time beyond the
outer diameter of the groove ($\sim$1\% of the total active volume). 
Interactions
in the rest of the detector volume result in the most common type I
trajectories.

\subsection{Discrimination between single--site and multiple--site events}
\label{sec:BEGeModelingPSD}
In the previous subsection we have shown that most of the single interactions in
BEGe detectors produce uniform signals (type I trajectories), 
differing one from the other only by
the total charge collection times. 
In addition, the current signals have a simple shape with only a single narrow
peak at the end of charge collection.
These features can be exploited for a powerful discrimination of Single--Site Events
(SSE) and Multiple--Site Events (MSE). A discrimination method based on the
current signal amplitude was introduced in Ref.~\cite{bib:dusan1} and explained
by using an empirical estimate of the weighting fields in
BEGe detectors~\cite{bib:dusan2}. Here we recall the basic idea 
of the method and refine the
discussion based on more accurate electric potential and field calculations.

\begin{figure}[t]
   \scriptsize
   \begin{center}
      \input{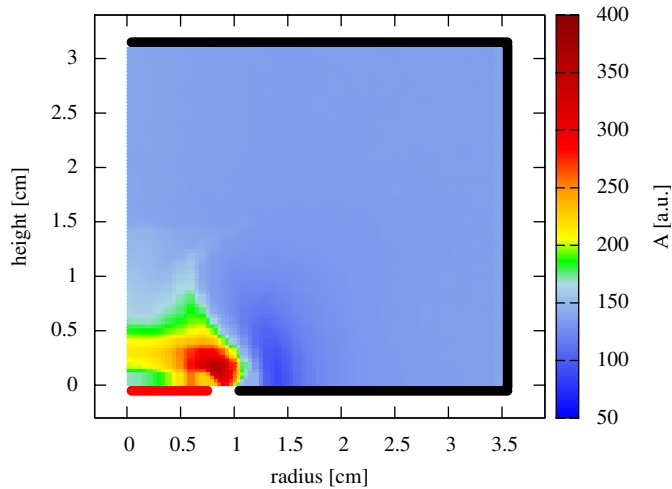}
   \end{center}
   \caption{
   Distribution of the maximum current pulse amplitude $A$ for
   simulated single interactions with unit energy as a function of the
   interaction point. $A$ is constant in most of the detector  volume (type I
   trajectories), but it is amplified in the region close to the cathode (type II
   trajectories).
   The region of lower $A$ values close to the outer radius of the groove is an
   artifact due to the use of the pulse shape library with $1~\mbox{mm}$ step size. Here
   the charge collection time varies on scales smaller than the library grid
   with the effect that the averaging of library pulses leads to a reduction in
   current--peak amplitude. The effect is not present when the signals are
   generated directly without the use of the library.
   }
   \label{fig:A}
\end{figure}

For typical BEGe events the part of the hole trajectory that passes through strong
$E_w$ is always the same. Therefore, according to equation~\eqref{eq:SRC} the amplitude of
the induced current signal depends only on the total charge of the considered
hole cluster. \figurename~\ref{fig:A} shows the value of the maximum current
pulse amplitude $A$ for simulated single interactions with unit energy
deposition, as a function of the interaction position in a vertical section of a BEGe
detector.
We see that the parameter $A$ is constant in most of the detector
volume (corresponding to the typical type I trajectories)
except for the region close to the p+ electrode, where the type II trajectories with
amplified current signals originate
(\figurename~\ref{fig:weightingTrajectories2}).

\begin{figure}[t]
   \scriptsize
   \begin{center}
      \input{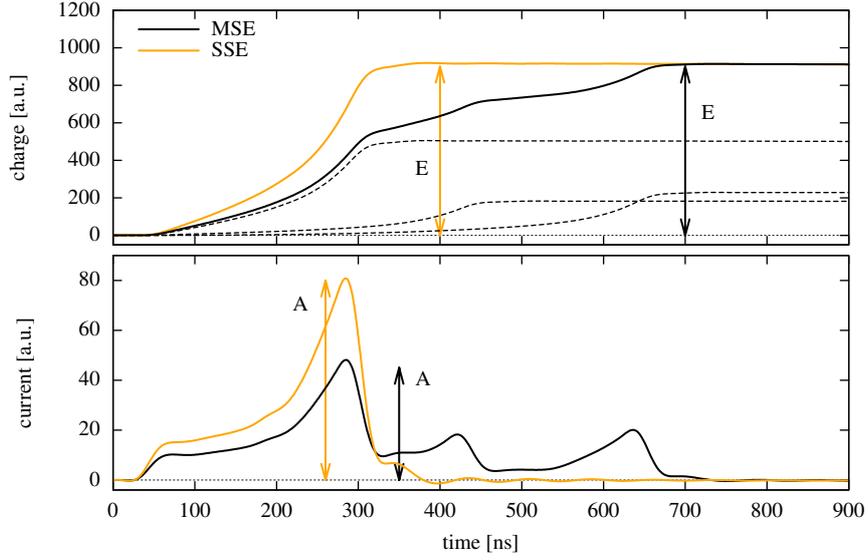}
   \end{center}
   \caption{The upper panel shows typical charge pulses for SSE and MSE
   while the lower panel shows the corresponding current pulses obtained as the
   derivatives of the charge signals. The dashed lines in the top panel show the contributions of
   the individual single interactions to the total charge pulse in the MSE.}
   \label{fig:PSA}
\end{figure}

The Pulse Shape Discrimination (PSD) method uses the parameter $A$ normalized to
the total event energy $E$: the $A/E$ ratio. The concept is depicted in
\figurename~\ref{fig:PSA}.
For SSE, all energy is transferred to a single charge
cluster -- it follows that $A/E$ is constant for such events. In MSE, the
total event energy is shared between several spatially--separated charge
clusters (\figurename~\ref{fig:PSA} illustrates the case of three clusters). Since the current peaks are narrow and the charge
collection time is position--dependent, the $A/E$ ratio will be smaller than the
constant value given by SSE.

The region of amplified current signals (approximately a hemisphere of a  $\sim 13~\mbox{mm}$ radius from the
center of the p+ electrode in \figurename~\ref{fig:A}, corresponding to $\sim 5\%$ of
the detector active volume) was already identified in
Refs.~\cite{bib:dusan1,bib:dusan2}. It limits the efficiency of our PSD method, since
interactions from MSE occurring in this volume can have their $A/E$ ratio
amplified above the SSE discrimination threshold. 
On the positive side, this effect can be used to identify surface events
occurring on the p+ electrode and groove surfaces.

\section{Validation of the simulation}\label{sec:validation}
The implementation of BEGe detectors in the \MaGe~Monte Carlo simulation
framework, accurately reproducing their radiation detection efficiency, was
presented already in \cite{bib:dusan1,bib:dusan2,bib:MatteoThesis,bib:assunta}.
Here we report only the measurements carried out to validate the pulse shape
simulation. Two sets of measurements were performed for this purpose. First we
used a collimated $^{241}\mbox{Am}$ source to generate well localized
interactions and to compare directly the pulse shapes for different interaction
positions close to the surface of the detector.  Then, a $^{228}\mbox{Th}$
source was used to investigate the distribution of the pulse rise times and of
the parameter $A/E$ as a function of the energy for events in the whole
detector volume.  
\subsection{The experimental setup}\label{sec:validationExpSetup}
The detector used for the validation measurements was the BE3830/s BEGe detector described in
Section~\ref{sec:BEGeModeling}. 
The front end read out of the signals was performed with the
Canberra charge sensitive preamplifier 2002CSL~\cite{bib:BEGe}, which is 
integrated in the housing of the detector. The preamplifier output was digitally
recorded with a 4 channel N1728B CAEN NIM flash  
analogue--digital converter~\cite{bib:tnt}
running at $100~\mbox{MHz}$ sampling frequency with a precision of 14 bits. 
This module is fitted with a USB connection for communication with a PC.
To control the digitizer setup, the acquisition parameters and the storage of the 
data, the PC was interfaced to the NIM module by using the TUC 
software~\cite{bib:TUC}.
We recorded pulse shapes with a total length of $40~\mu\mbox{s}$ including a
baseline before the signal of $\sim 10~\mu\mbox{s}$.
The energy reconstruction was performed off--line by using the 
\GELATIO~software~\cite{bib:gelatio} implementing the \emph{Moving Window
Deconvolution} approach~\cite{bib:gast} and several digital filters were applied to the data
to reject pile--up and noise events.

To perform the $^{241}\mbox{Am}$ measurements, 
a mechanical device was built to allow the movement of 
collimated sources with sufficient accuracy along the diameter of the 
front face and circularly around the symmetry axis of the detector. The
collimator had a hole of $1~\mbox{mm}$ diameter and a length of
$34~\mbox{mm}$. 
These $^{241}\mbox{Am}$ collimated measurements were accurately simulated by using a
conical beam of $59.5~\mbox{keV}$ photons.
\subsection{Pulse shape comparison with low energy gamma--ray beams}
\label{sec:validationPSC}
Using a collimated beam of low energy gamma--rays ($59.5~\mbox{keV}$) from a
$^{241}\mbox{Am}$ source allows to obtain pulse shape data with well defined
interaction coordinates. As the $59.5~\mbox{keV}$ photons penetrate only a few
millimeters underneath the germanium crystal surface and the volume spread of their
energy deposition is of similar size, the narrow beam can be used to study
pulse shape position dependence with few millimeters resolution. As the
topologies of the individual photon interactions are statistically variable,
we selected events with energy deposition corresponding to the $59.5~\mbox{keV}$
peak and calculated the average signals for each position of the collimated
source.
This averaging procedure also reduced significantly the electronic noise.

The pulse shape comparison included scanning along the diameter of the detector
and a circular scan at a fixed distance from the symmetry axis of the detector.
The radial scan allows the study of pulse shapes as a function of the distance 
from the center. With the circular scan the rise time 
variations due to the drift velocity anisotropy caused by the crystal lattice
structure were investigated.

\begin{figure}[tbh]
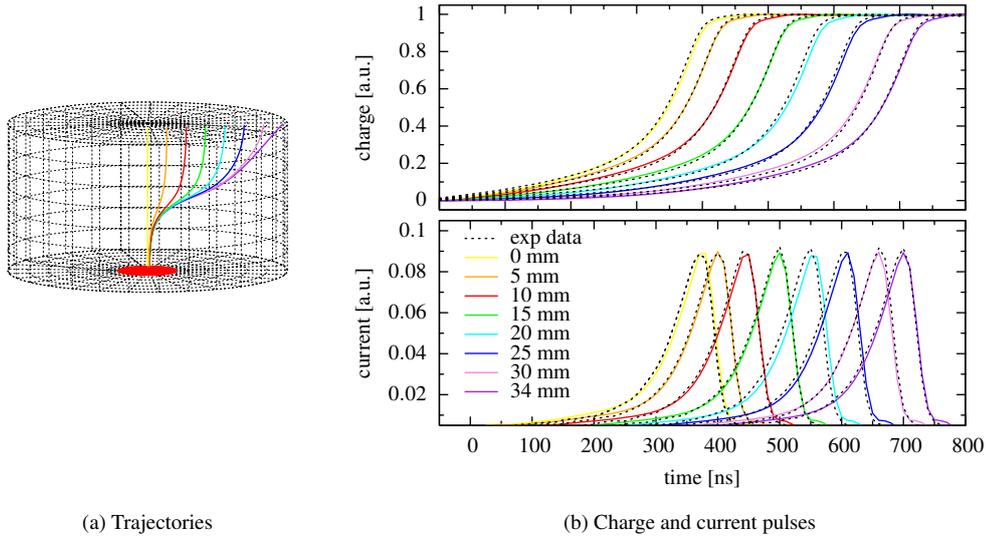

  \begin{center}
  {
    \scriptsize
      \input{validationTrajectories}
      \put(-80,-15){\makebox(0,0)[r]{\strut{}(a) Trajectories}}%
      \input{validationRadialPulses}
      \put(-80,-15){\makebox(0,0)[r]{\strut{} (b) Charge and current pulses}}%
  } 
  \end{center}
  \caption{%
  Simulated hole trajectories (a) and average signals (b) computed
  for several interaction points at different radii on the detector top
  surface.
  The interactions were obtained by moving a $1~\mbox{mm}$ collimated
  $^{241}\mbox{Am}$ source on the top surface of the end--cap along 
  the radial direction.
  The small deviation present at the very beginning of the signals can be related 
  to the fixed grid width used for defining the library of pulses.
  }
  \label{fig:radialscan}
\end{figure}

The radial scan was performed by moving the collimated 
$^{241}\mbox{Am}$ source along the radius of the detector on the top surface of
the end--cap in steps of $0.5~\mbox{cm}$.
The simulated hole trajectories for each collimator position and the average
signals are shown in \figurename~\ref{fig:radialscan}. 
Good agreement is observed between the simulated and experimental pulses.

\begin{figure}[t]
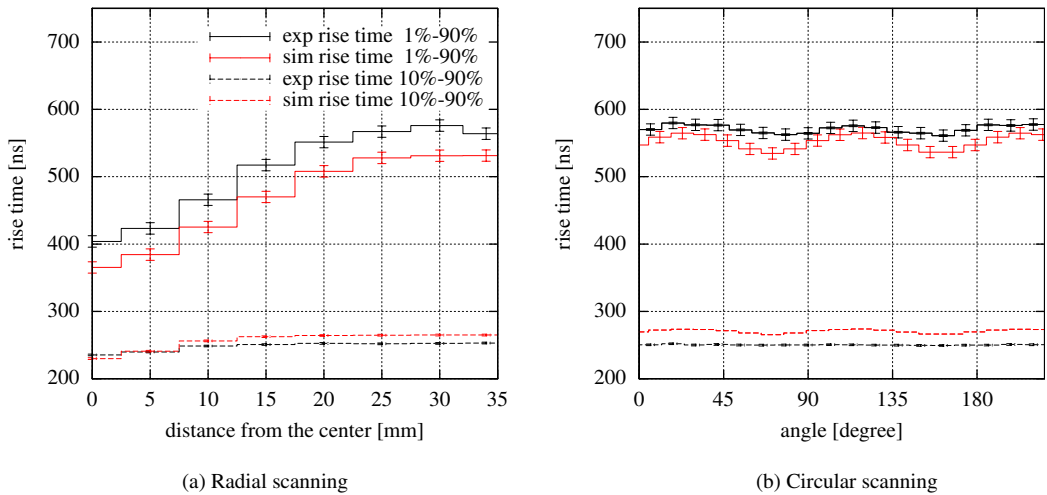

  \scriptsize
  \begin{center}
  { 
     \input{validationRadialScanning}
      \put(-80,-15){\makebox(0,0)[r]{\strut{} (a) Radial scanning}}%
     \input{validationCyrcularScan}
      \put(-40,-15){\makebox(0,0)[r]{\strut{} (b) Circular scanning}}%
  }
  \end{center}
  \caption{%
    $1\%-90\%$ and $10\%-90\%$ rise time values reconstructed
    from the simulated and experimental average pulses for different interaction
    points.
    The interactions were obtained by moving a $1~\mbox{mm}$ collimated
    $^{241}\mbox{Am}$ source on the top surface of the end--cap, along
    the radial direction (a) and around a part of a circle centered on the detector symmetry
    axis (b).
    The rise time uncertainty was estimated by determining the spread of
    rise time values computed from several measurements at one source position.
    }
  \label{fig:risetimeScan}
\end{figure}

As an increase of the radius value corresponds to an increment of the
hole collection time, the total rise time should increase as the interaction
point moves away from the symmetry axis. However, according to the discussion in 
Section~\ref{sec:BEGeModelingWeightingField}, for interactions far from the p+ electrode the pulse shapes are
expected to be different only at the very beginning. Looking at
\figurename~\ref{fig:risetime}.b BEGe signals along the top surface show only minimal variation of the
$10\%-90\%$ rise time. In order to observe a significant
effect, the rise time between $1\%$ and $90\%$  has to
be considered. This can be clearly seen in \figurename~\ref{fig:risetimeScan}.a for both
the experimental and simulated data.

For the circular scan the collimated $^{241}\mbox{Am}$ source was moved around a
circle with a radius of $34~\mbox{mm}$ centered on the detector's
symmetry axis on the top surface of the end--cap with a  $\sim10^{\circ}$ step.
The drift velocity anisotropy arises due to the charge carriers drifting at
different angles relative to the detector's crystallographic axes. Like with the
radial scanning, also in this case the differences occur only in the first part
of the charge collection, and the $1\%-90\%$ rise time has to be
used. The comparison of circular--scan rise times as a function of the
circumference angle from the simulation and the experiment is shown in
\figurename~\ref{fig:risetimeScan}.b.

Although the experimental data show a behavior coherent with the simulation in
both the radial and circular scans, the agreement is only qualitative.
The $1\%-90\%$ rise times of the simulated data are systematically shifted by
$20-40~\mbox{ns}$ ($5\%-10\%$ of the total rise time) to lower values than the
experimental data.
In contrast, the $10\%-90\%$ rise times of the simulated data are shifted by 
$\sim20~\mbox{ns}$ to higher values than the experimental data.
The change in the sign of the discrepancy is due to differences in the shape of
the average pulses.
Furthermore, the oscillation amplitude due to crystal
anisotropy in \figurename~\ref{fig:risetimeScan}.b is higher in the simulation
than in the experimental data by
$30\%$. Unlike the experimental data, the simulation shows a residual oscillation of
about $5~\mbox{ns}$ also in the $10\%-90\%$ rise time plot. From these comparisons we can
conclude that the simulation is more sensitive to the drift velocity anisotropy
than the real detector and probably further small discrepancies in the signal
calculation are present. This allows us to identify several areas of possible
improvements of our simulation, including the accuracy in the geometry
definition (especially important in case of the p+ electrode), the finite grid
width used for defining the library of pulses and the charge carrier mobility
model parameterizations (defined by using a different
crystal~\cite{bib:bruynell01}). Also further measurements could be performed to
characterize the position dependence of the BEGe signals (e.g. by
performing collimated $^{241}\mbox{Am}$ scans also along side and back surfaces,
and by doing collimated single--Compton scattering measurements with
higher--energy sources to sample also the internal detector volume).
\subsection{Rise time and $A/E$ distributions studies}
\label{sec:validationIntegrate}
A different approach for validating the simulation is given by the comparison of
the rise time and the $A/E$ distributions as a function of the event energy.
Unlike the $^{241}\mbox{Am}$ measurements which create interactions restricted to a region close to the
detector surface, $2.6~\mbox{MeV}$ gamma--rays from a $^{228}\mbox{Th}$ source interact in the whole
detector volume and can deposit energy at several sites before escaping or
being absorbed. Under these circumstances, the distributions of the rise times
and of the parameter $A/E$ are sensitive to the electric and weighting potential in
the whole detector and thus are useful for testing the accuracy of the
simulation.

For this study a long measurement was performed with a $^{228}\mbox{Th}$
source\footnote{In the simulation only the $^{208}$Tl isotope was considered
resulting in some missing gamma--ray interactions at energies below $\sim
1.8~\mbox{MeV}$ as compared to the experimentally measured spectrum. 
Moreover, the simulation does not include the SEP Doppler
broadening of the positron annihilation energy which is determined by the
electron momentum distribution in germanium.}. 
The detector was located in the
LNGS underground laboratory and surrounded by a lead shielding in order to reduce
the background due to cosmic rays and natural radioactivity.

\begin{figure}[t]
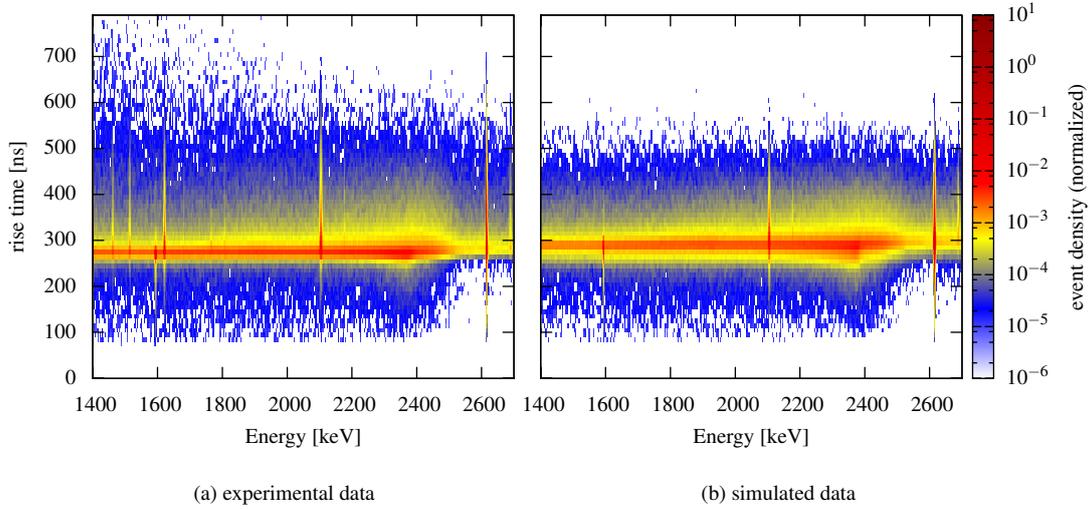

   \begin{center}
      \scriptsize
      \hspace{-0.5cm}
      \input{drt_exp} \hspace{-2.5cm}
      \input{drt_sim}
      \put(-260,0){\makebox(0,0)[r]{\strut{} (a) experimental data}}%
      \put(-80,0){\makebox(0,0)[r]{\strut{} (b) simulated data}}%
   \end{center}
   \caption{Experimental and simulated distributions of the $10\%-90\%$ rise time values as a
   function of energy from the $^{228}\mbox{Th}$ measurements. In the simulation only 
   $^{208}$Tl is considered. 
   The density plots are normalized according to the number of events in the
   $2.6~\mbox{MeV}$ FEP.
   }
   \label{fig:rt_distribution}
\end{figure}

\figurename~\ref{fig:rt_distribution} shows the distribution of the $10\%-90\%$ rise time
as a function of the energy in the range including the Double Escape Peak (DEP)
at $1592.5~\mbox{keV}$ and the Full Energy absorption Peak (FEP) at
$2614.5~\mbox{keV}$ of $^{208}\mbox{Tl}$. 
Both the experimental and simulated data show a similar structure of the
distribution, with a high density band at rise time values around $\sim
270~\mbox{ns}$.  The region below the band contains the fast pulses resulting
from events with type II charge collection trajectories described in
Section~\ref{sec:BEGeModelingWeightingField} generated by interactions close to
the p+ electrode.  The region above the band is composed mainly of MSE events
which have typically slower rise times than SSE (this is visible in
\figurename~\ref{fig:PSA}). Consistently, all the full absorption peaks
(consisting typically of multiple Compton scatterings followed by
photo--absorption) show an important tail in this region, while the DEP
(containing the typically single--site e$^-$ and e$^+$ absorptions after
pair--production) has a very weak tail. According to the simulation, events
occurring in a small volume close to the outer radius of the groove (type III
charge collection trajectories in Section~\ref{sec:BEGeModelingWeightingField})
are also expected to have increased rise times. It is however evident from the
plot that the simulation creates events with rise times only up to
$600~\mbox{ns}$ while the experimental data show a significant number of events
with rise times $>600~\mbox{ns}$ ($\sim 1.5\%$ of the total).

\begin{figure}[t]
   \scriptsize
   \begin{center}
      \input{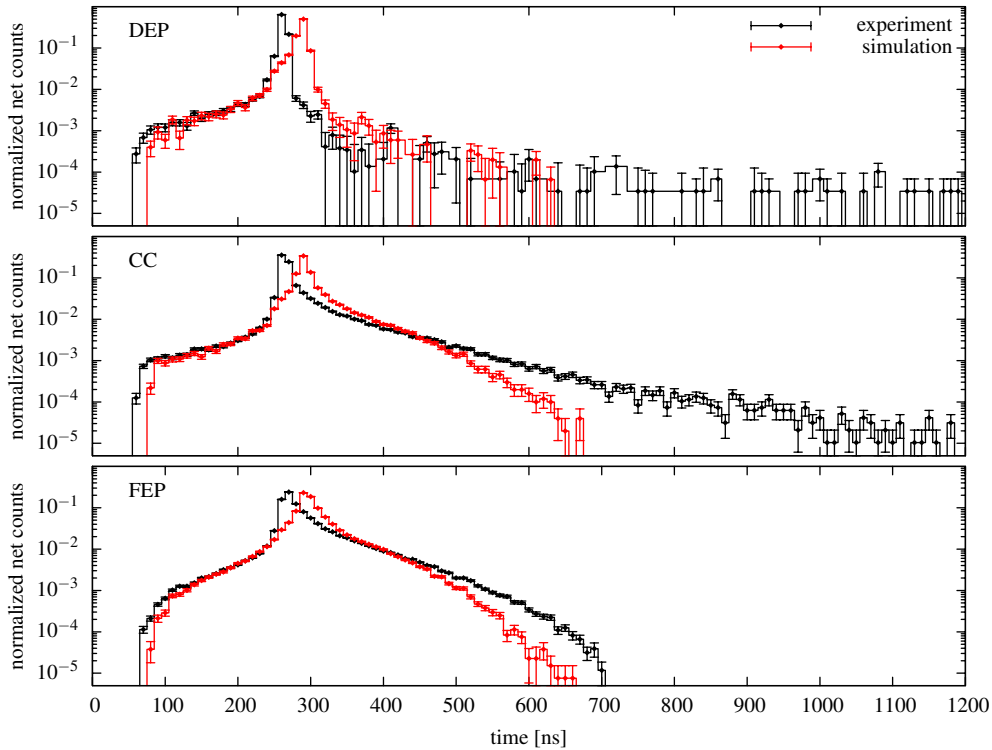}
   \end{center}
   \caption{%
   Experimental and simulated distributions of the rise time for the
   region around the DEP, the Compton continuum at $2~\mbox{MeV}$ (CC) and the
   $2.6~\mbox{MeV}$ FEP. The error bars take into
   account only the statistical error. The DEP and the FEP distribution are
   corrected for the background.
   The integrals of the histograms are normalized to one.
   }
   \label{fig:rt_distribution_slides}
\end{figure}

The simulated and experimental data can be compared in more detail in
\figurename~\ref{fig:rt_distribution_slides} which shows distributions of the
$10\%-90\%$ rise time in narrow energy regions around the DEP, the Compton continuum at
$2~\mbox{MeV}$ and the $2.6~\mbox{MeV}$ FEP.  
For the DEP and FEP distributions the contribution from their
Compton continuum backgrounds is subtracted. While the shape of the
distribution below $\sim 250~\mbox{ns}$ (corresponding to the fast signals from
the type II events) is reproduced fairly well, the main peak and the region
above it show some differences between the simulation and experimental
histograms. As we already noted in the previous subsection, our simulation is
not perfect in some aspects. The charge carrier mobility model is
likely responsible for the shift of the rise time histogram maximum, while the
enhanced drift velocity anisotropy in the simulation could explain the greater
width of the peak. Minor discrepancies in the geometry description of the p+
electrode and the groove, as well as the finite grid width of our pulse library
could account for the small $\sim 20~\mbox{ns}$ shift between the start of the
experimental and simulated histogram, as well as for a part of the excess long
rise time events in the experiment. In addition, some pile--up events show up
with very long rise times up to a few $\mu \mbox{s}$.

However, most of the signals with rise time above $\sim 500~\mbox{ns}$ (most
clearly visible in the Compton continuum histogram in
\figurename~\ref{fig:rt_distribution_slides}) are expected to come from 
the region close to the n+ surface. The Li--diffused n--side of the p--n junction
can not be fully depleted of conduction band electrons and forms the well known
dead layer covering the outer surfaces of p--type Ge detectors.  
The appearance of pulses  with long rise times in n+ surface interactions 
was noticed before with conversion electron measurements in Ref.~\cite{bib:koln} and can be clearly observed by
irradiating the detector with $^{241}\mbox{Am}$ gamma rays, as reported
in Ref.~\cite{bib:slowPulses}. 
It can be assumed that the electric field near the n+ surface has
insufficient strength to effectively move the charge carriers.
In turn, the drift time is extended and part of the
charge can be lost by trapping.
This effect can explain the observed slow pulses
in our experimental data. It is also consistent with the fact that the DEP and
FEP peak regions in \figurename~\ref{fig:rt_distribution_slides} show a much smaller
excess of slow pulses than the Compton continuum region. This is due to the
charge carrier losses moving the event in the energy spectrum
away from the peak regions into the lower energy continuum. As the peak histograms
have the continuum background subtracted, the slow pulse
contribution is removed as well.
The content of these slow pulses in the Compton continuum region was estimated
to be $\lesssim4\%$.
The study of these n+ layer effects can be in future augmented by our pulse shape
simulation if the spatial resolution of the field calculation is improved to
better cover the $0.7~\mbox{mm}$ thick region.

\begin{figure}[t]
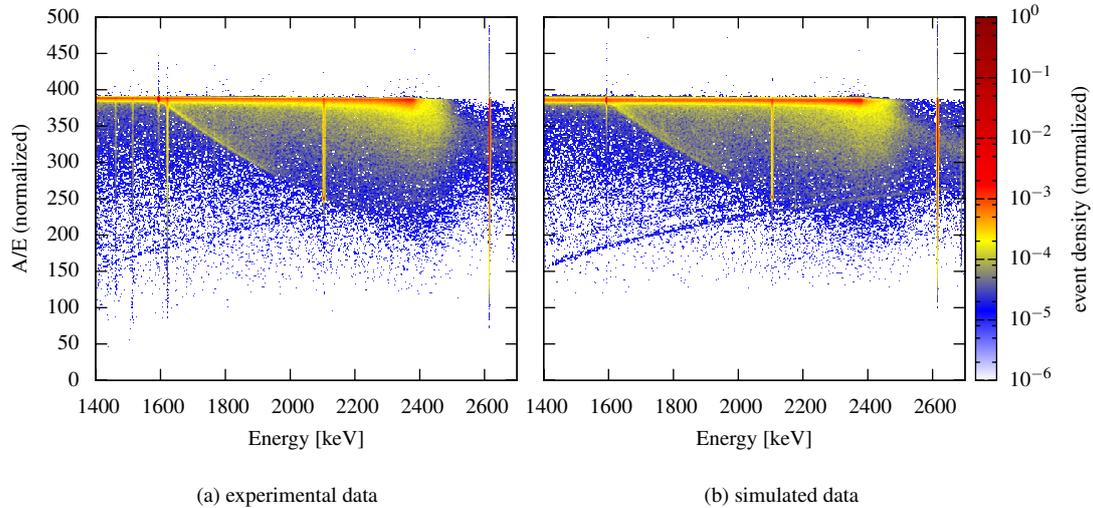

   \begin{center}
      \scriptsize
      \hspace{-0.5cm}
      \input{dae_exp} \hspace{-2.5cm}
      \input{dae_sim} 
      \put(-260,0){\makebox(0,0)[r]{\strut{} (a) experimental data}}%
      \put(-80,0){\makebox(0,0)[r]{\strut{} (b) simulated data}}%
   \end{center}
   \caption{Experimental and simulated distributions of the $A/E$ parameter as a
   function of energy for $^{228}\mbox{Th}$ source measurement. In the simulation only
   the $^{208}\mbox{Tl}$ is considered.
   The density plots are normalized according to the number of events in the
   $2.6~\mbox{MeV}$ FEP.
   }
   \label{fig:ae_distribution}
\end{figure}

\figurename~\ref{fig:ae_distribution} shows the distribution of the parameter
$A/E$ as a function of energy.
The plots show a well defined horizontal band of increased event density,
composed of SSE. Below the band the region of MSE extends. Above the band one
can see a region of events with amplified current signal occurring close to the
p+ electrode as already discussed in Section~\ref{sec:BEGeModelingPSD}.  This
interpretation of the $A/E$ distribution was previously validated by coincident
Compton scattering measurements (providing clean SSE samples at several
energies) and by collimated $^{228}\mbox{Th}$ measurements (providing DEP events
restricted to the detector volume close and far from the p+
electrode)~\cite{bib:dusan1,bib:dusan2}.  Another visible feature in the plot is a
diagonal band between the DEP and the Single
Escape Peak (SEP), composed of pair--production events
(as explained in Ref.~\cite{bib:dusan1,bib:dusan2}), and two weak narrow bands
in the MSE region, which are caused by cascade summing events\footnote{The two
visible additional bands result from the summation of the SSE in the Compton
continuum with either the $511~\mbox{keV}$ or $583~\mbox{keV}$ full energy gamma absorption from
the $^{208}$Tl cascade. This results in the energy of the event being shifted but
the $A$ value staying the same since the SSE from the Compton scattering still
dominates the current signal.}. 
There is no significant difference between the plots from the
experimental and simulated data apart from the $^{212}\mbox{Bi}$ lines present in the
measurement and not in the simulation. All the features of the $A/E$ distribution
are reproduced by the simulation.

\begin{figure}[htb]
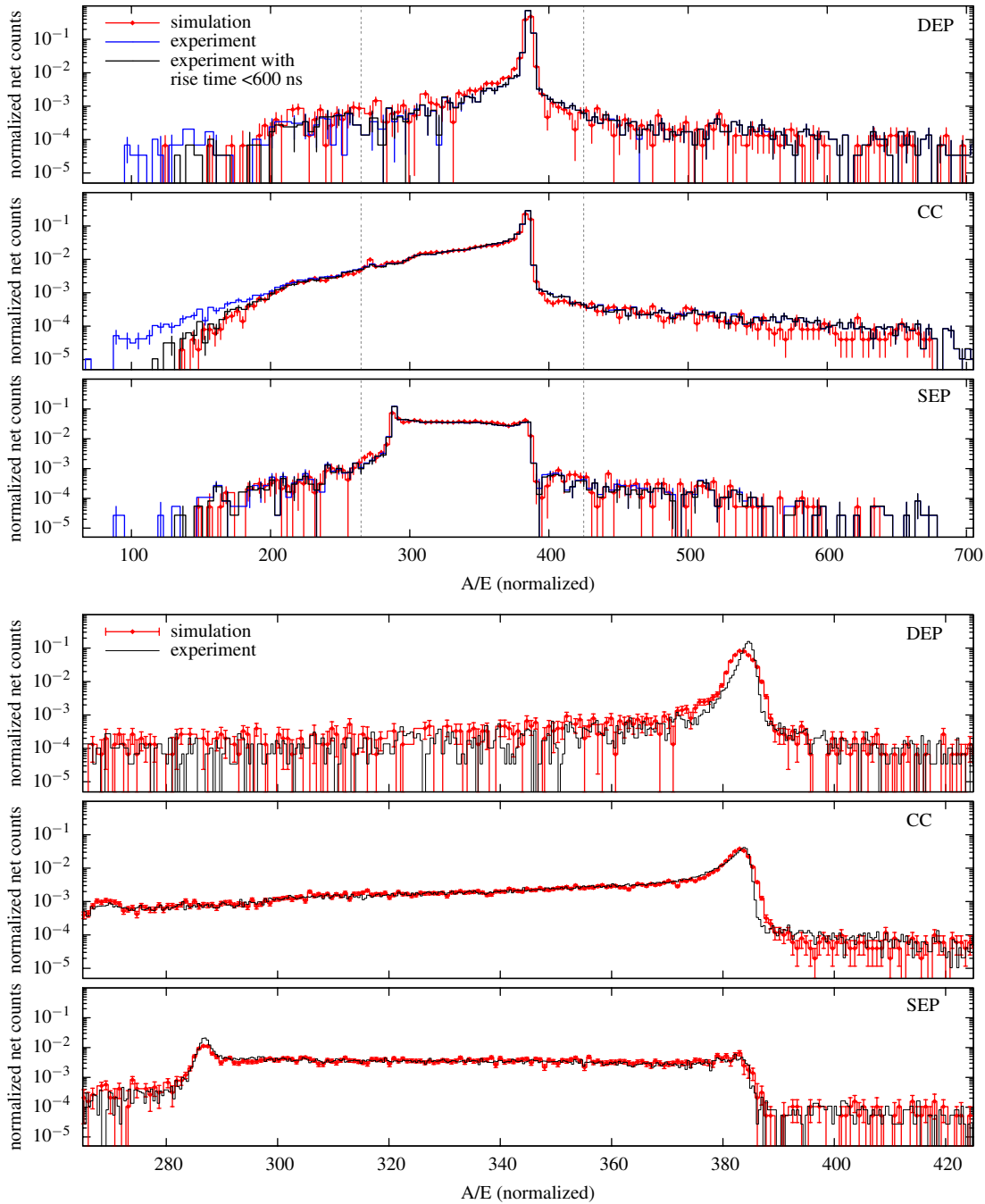

   \scriptsize
   \begin{center}
      \input{validationAEdistribution.tex}
      \input{validationAEdistributionZoom.tex}
   \end{center}
   \vspace{-0.5cm}
   \caption{%
   Experimental and simulated distribution of the $A/E$ parameter
   in the region around the DEP, the Compton continuum at $2~\mbox{MeV}$ (CC) and the SEP. 
   The error bars take into account only the statistical error. The DEP and the SEP distributions are
   corrected for the background.
   The integrals of the histograms are normalized to one.
   The top plots show the experimental distribution before and after a
   cut on the rise time to remove slow pulses not included in the simulation.
   The regions defined by the gray vertical lines are enlarged in the bottom plots.
   }
   \label{fig:ae_distribution_slides}
\end{figure}

\figurename~\ref{fig:ae_distribution_slides} shows the $A/E$ distribution for
energy slices around the DEP, the Compton continuum at $2~\mbox{MeV}$ and the
SEP at $2103.5~\mbox{keV}$.
As in \figurename~\ref{fig:rt_distribution_slides} the histograms from the DEP
and SEP are corrected for the Compton continuum background contribution. The
simulation is in good agreement with the experimental data in all three energy
regions.  However, in the Compton continuum region one can see a small deviation at low $A/E$
values.
This excess can be associated to the presence of the near n+ electrode slow pulses in experimental data. 
It can be assumed that by stretching the charge carrier cluster, the peak in the current
signal gets wider and its height is reduced. 
By applying a cut on the rise time at $600~\mbox{ns}$ (black histogram in
\figurename~\ref{fig:ae_distribution_slides}) the agreement in the tails improves
significantly. 

The plots at the bottom of \figurename~\ref{fig:ae_distribution_slides} show the
$A/E$ histograms of Compton continuum, the SEP and the DEP focused on the region
near the SSE--band peak (near $A/E\sim385$).
It is apparent mainly from the DEP plot that the simulated SSE peak is a little wider
than in the experimental data and the shape is slightly different. This
discrepancy is likely the result of the same simulation inaccuracies as
discussed earlier. 
The Compton and SEP histograms show that for MSE--dominated
distributions the overall agreement is significantly better. This demonstrates
that the partitioning of the energy depositions into individual charge clusters
(performed by the Monte Carlo block) is simulated accurately.

\section{Background rejection and signal identification studies}
A direct application of the BEGe detector simulation was the
study of the Pulse Shape Discrimination (PSD) efficiency in separating SSE
from MSE introduced in the Section~\ref{sec:BEGeModelingPSD}.
First we compared the performances of the PSD applied
to experimental and simulated sets of measurements with $^{228}\mbox{Th}$
and $^{60}\mbox{Co}$ sources placed outside the detector housing.
We then used the simulation to study the survival probabilities of decays which
occur inside the Ge crystal. The focus of this investigation concerned the
\nubb~decay signal of $^{76}\mbox{Ge}$ and the background signals created by the
decays of the radio-isotopes
$^{68}\mbox{Ge}$ and $^{60}\mbox{Co}$. 
The latter are produced by cosmic ray interactions during
germanium detector production above ground. 
These radio-isotopes are one of the
most critical backgrounds in $^{76}\mbox{Ge}$ neutrinoless double beta decay
search.
However it is challenging to study experimentally the decays of these isotopes inside
the Ge crystal.
The simulation tools introduced in this paper represent a powerful method to
investigate quantitatively the PSD performance on these backgrounds.

\subsection{Comparison of pulse shape discrimination of simulated and experimental data}\label{sec:calibration}
The comparison was performed using two high--statistic data sets generated
by a $^{228}\mbox{Th}$ source ($^{208}\mbox{Tl}$ in the simulation) placed
$2~\mbox{cm}$ away from the Al end--cap and a
$^{60}\mbox{Co}$ source placed at the end--cap (on the detector symmetry axis).
The recorded signals ($10~\mbox{ns}$ time step) are smoothed by triple
$50~\mbox{ns}$ moving window averaging. 
The current signal is then reconstructed by $10~\mbox{ns}$ differentiation of
the smoothed charge pulse.
Finally, interpolation into $1~\mbox{ns}$ time steps is performed before
measuring the current pulse amplitude $A$.
This procedure was found to provide the optimal resolution for $A$.

The SSE--rich Compton continuum of the $2.6~\mbox{MeV}$ $^{208}\mbox{Tl}$ gamma--line at
$2.6~\mbox{MeV}$ is
used to calibrate the slight energy dependence of the SSE band in the $A/E$
distribution. This procedure was described in detail and validated in
Refs.~\cite{bib:dusan1,bib:dusan2}.  The multi--site events, which have lower $A/E$ values
than single--site events, are removed by performing the PSD cut at a constant offset from the
center of the SSE band (visible as a high--density band at $A/E\sim385$ in
\figurename~\ref{fig:ae_distribution}).  

The efficiency of the cut is tuned by the definition of the offset.
In Refs. \cite{bib:dusan1,bib:dusan2}, the cut offset was defined according to the
spread of the SSE band in the DEP region and the resulting DEP acceptance was close to $90\%$.
However, using this definition of the PSD cut on the simulated data is expected
to create a biased result, since the profile of the SSE--band in our simulation
has a different shape and spread than the one extracted from the experimental
data (see \figurename~\ref{fig:ae_distribution_slides} and discussion in the
previous section). On the other hand, the profile of the multi--site event region is
dominated by the partitioning of the charge clusters, so the overestimation of
anisotropy or other limitations in the simulation have a relatively smaller effect on the
MSE as compared to the SSE distribution (see \figurename~\ref{fig:ae_distribution_slides}).  

\begin{figure}[t]
   \scriptsize
   \begin{center}
      \input{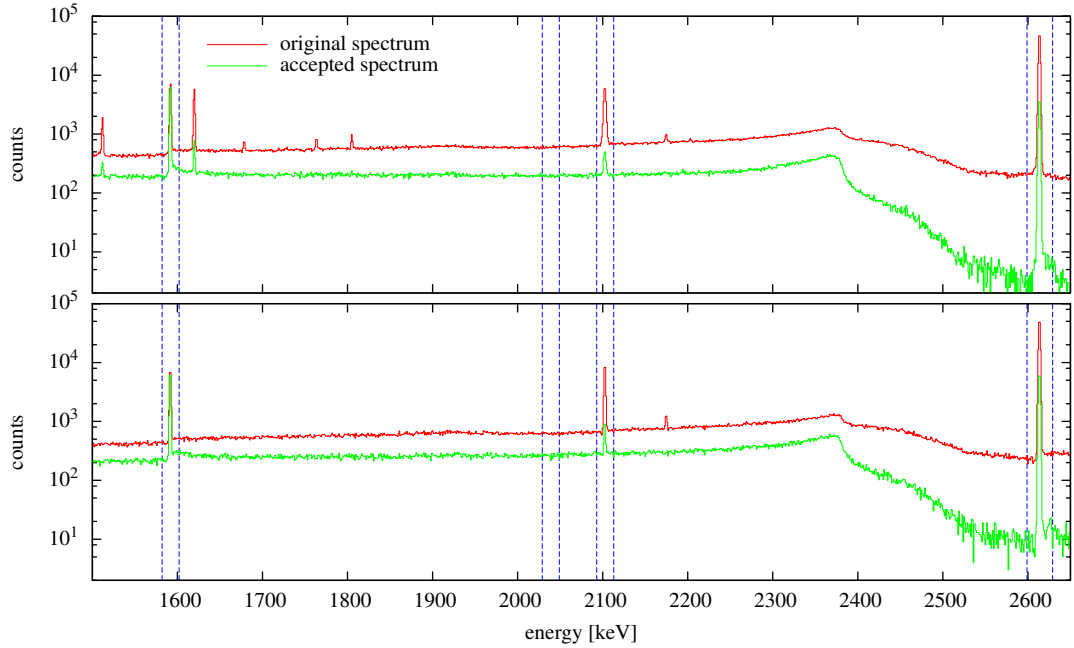}
   \end{center}
   \caption{Experimental (top) and simulated (bottom) energy spectra for a Th
   source before (red line) and after (green line) applying the PSD cut.}
   \label{fig:psdExt}
\end{figure}

To avoid the effect of the SSE--band discrepancy in the simulation results, we developed alternative definitions of the $A/E$ cut offset.
For experimental data we apply cut offset that retains a fixed acceptance of the
net area of the DEP at $90\%$, maintaining compatibility with the previously
published results. The survival fractions for $^{228}\mbox{Th}$ and
$^{60}\mbox{Co}$ are listed in the ``experimental'' column of
\tablename~\ref{tab:psdExt}. 
\begin{table}
   \caption{%
   Percentage of survived events from a $^{228}\mbox{Th}$ and a
   $^{60}\mbox{Co}$ measurement after the PSD cut in different regions of
   interest (ROI) for the three cut definitions. The results
   for DEP, SEP, FEP, SP and the two $^{60}\mbox{Co}$ lines at
   $1173~\mbox{keV}$ and $1332~\mbox{keV}$ (FEP1 and FEP2 in the table) 
   are calculated using the net peak areas. 
   The CC region corresponds to an $80~\mbox{keV}$ wide section of Compton
   continuum centered at $2039~\mbox{keV}$ ($^{76}\mbox{Ge}$ \Qbb). The
   uncertainties given in parenthesis for the least significant digits 
   include statistical as well as systematic uncertainties estimated by cut parameter variation. 
   The $^{208}\mbox{Tl}$ FEP and the $^{60}\mbox{Co}$ FEP1 lie outside of the
   range covered by the calibration of $A/E$ energy dependence ($1.35~\mbox{MeV}$
   to $2.38~\mbox{MeV}$) so can be subject to additional systematic errors.
   }
   \begin{center}
   \renewcommand{\arraystretch}{1.1}
   \begin{tabular}
      {| c | c | 
      p{0pt} l @{ (}  c @{)} p{10pt}|
      p{0pt} l @{ (}  c @{)} p{10pt} 
      p{0pt} l @{ (}  c @{)} p{10pt} 
      p{0pt} l @{ (}  c @{)} p{10pt}| }
      \cline{3-18}
      \multicolumn{2}{c}{}
      & \multicolumn{4}{|c}{experimental} 
      & \multicolumn{12}{|c|}{simulated} \\
      \hline
      \multicolumn{1}{|c}{source} 
      & \multicolumn{1}{|c|}{ROI} 
      & \multicolumn{4}{c|}{DEP} 
      & \multicolumn{4}{c}{DEP} 
      & \multicolumn{4}{c}{SEP} 
      & \multicolumn{4}{c|}{SP }\\
      \multicolumn{1}{|c|}{}
      & \multicolumn{1}{|c|}{}
      & \multicolumn{4}{c|}{fixed to 90\%} 
      & \multicolumn{4}{c}{fixed to 90\%} 
      & \multicolumn{4}{c}{fixed to 5.45\%} 
      & \multicolumn{4}{c|}{fixed to 0.08\%}\\
      \hline
      $^{208}\mbox{Tl}$ & DEP    && 0.900   & 11 &&& 0.900  & 14  &&&  0.84   & 3  &&&   0.61   & 15 &\\
      $^{208}\mbox{Tl}$ & SEP    && 0.055   & 6  &&& 0.079  & 15  &&&  0.055  & 3  &&&   0.038  & 11 &\\
      $^{208}\mbox{Tl}$ & FEP    && 0.073   & 4  &&& 0.12   & 2   &&&  0.088  & 5  &&&   0.059  & 17 &\\
      $^{208}\mbox{Tl}$ & CC     && 0.341   & 14 &&& 0.42   & 3   &&&  0.357  & 17 &&&   0.25   & 7  &\\
      $^{60}\mbox{Co}$  & FEP1   && 0.113   & 6  &&& 0.138  & 17  &&&  0.105  & 6  &&&   0.07   & 2  &\\
      $^{60}\mbox{Co}$  & FEP2   && 0.106   & 6  &&& 0.133  & 17  &&&  0.102  & 6  &&&   0.07   & 2  &\\
      $^{60}\mbox{Co}$  & SP     && 0.00080 & 16 &&& 0.0021 & 7   &&&  0.0012 & 3  &&&   0.0008 & 3  &\\
      $^{60}\mbox{Co}$  & CC     && 0.0082  & 7  &&& 0.012  & 3   &&&  0.0073 & 9  &&&   0.0043 & 16 &\\
      \hline
   \end{tabular}
   \end{center}
   \label{tab:psdExt}
\end{table}
The results are very similar to those obtained in previous works -- the small
differences as compared to \cite{bib:dusan1} are mainly due to data
treatment\footnote{The different size of the detector used in \cite{bib:dusan1}
and geometry of its electrode was found to cause no visible difference in the
results if the same data treatment is applied. Only the result for the Compton
continuum region is significantly different, due to a larger distance of the
source from the detector, resulting in fewer cascade summing MSE.} and the
results in \cite{bib:assunta} differ due to a different implementation of the
analysis.  To be able to employ the simulation to estimate a realistic
experimental survival probability both in SSE--dominated and MSE--dominated
data, three different versions of the PSD cut were applied to the simulated
data.
The cut $A/E$ offsets were defined by fixing survival fractions in a different
peak region for each of the three cuts: 
1) in the DEP region; 
2) in the MSE--dominated SEP of $^{208}\mbox{Tl}$;
3) in the highly--MSE dominated Summation Peak (SP) of the two $^{60}\mbox{Co}$
gamma--lines.
The $A/E$ offsets of these cuts is
tuned to be equivalent exactly to the experimental data cut with $90\%$
acceptance of DEP events, i.e. the SEP--based cut is adjusted to $5.45\%$ SEP
survival and SP--based cut to $0.08\%$ SP survival (see the DEP, SEP and SP
entries in the ``experimental'' column of \tablename~\ref{tab:psdExt}).  The
results are shown in the last three columns of \tablename~\ref{tab:psdExt}. It
can be seen that the agreement with the experimental data for Compton
continuum\footnote{Part of the discrepancy in the $^{208}\mbox{Tl}$ Compton
continuum region arises due to the near n+ layer slow pulses, not included in
the simulation.} and the MSE dominated FEP and SP regions significantly improves
when the SEP--based cut is used as compared to the DEP--based cut, while the DEP
survival is accordingly underestimated. The SP cut results suffer from a higher
uncertainty due to the very small number of events remaining in the SP after
cut, nevertheless the agreement with experimental data is still reasonable in
the highly MSE--dominated regions. We can conclude that the more similar is the
event topology between the data region used for cut calibration and the studied
region, the better is the agreement between simulated and experimental results.
\figurename~\ref{fig:psdExt} shows the simulated and experimental
$^{208}\mbox{Tl}$ spectra before and after applying the PSD cut, using the
$90\%$ DEP cut definition.

\subsection{Survival fractions of decays internal the Ge crystal}\label{sec:psd}
\begin{figure}[tbp]
    \scriptsize
    \begin{center}
       \input{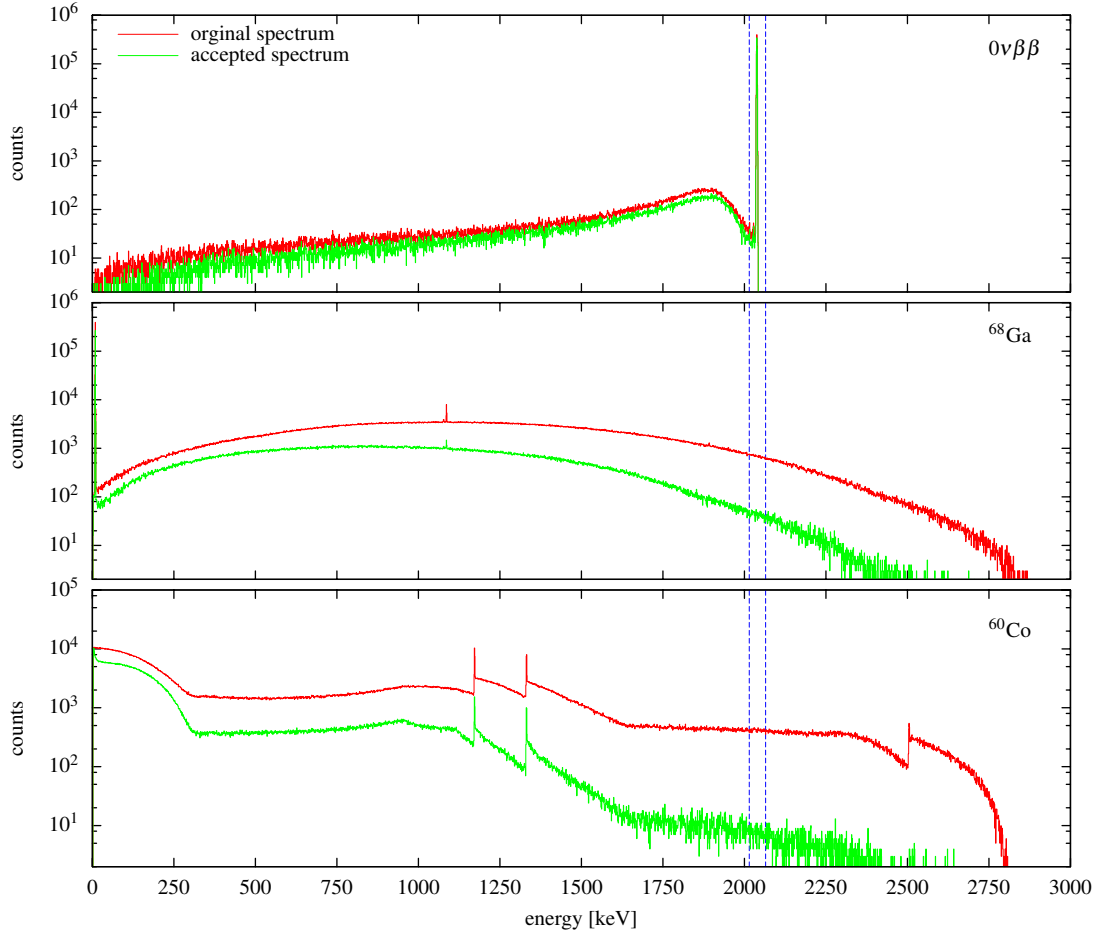}
    \end{center}
    \caption{Simulated spectra of \nubb~decays (top), $^{68}\mbox{Ga}$ (middle) and
    $^{60}\mbox{Co}$ (bottom) inside a BEGe detector before (red line) and after
    (green line) the PSD cuts.}
    \label{fig:internalBackground}
\end{figure}

In the last part of our study we applied the PSD to simulated $^{60}\mbox{Co}$,
$^{68}\mbox{Ga}$ decays and \nubb~decays of $^{76}\mbox{Ge}$, distributed inside the
Ge--detector volume. The aim was to evaluate the \nubb--signal acceptance by our
PSD method and to determine the suppression power of the intrinsic cosmogenic
backgrounds.  \figurename~\ref{fig:internalBackground} shows the spectra of
internal \nubb, $^{60}\mbox{Co}$ and $^{68}\mbox{Ga}$ before and after the PSD
cut using the DEP--based cut definition.
The resulting survival fractions in the energy regions around the
$^{76}\mbox{Ge}$ \Qbb~are summarized in \tablename~\ref{tab:psdInt},
including the DEP--based, SEP--based and SP--based PSD cut definitions.
According to the comparison reported in the previous section, we assume that the
most realistic value for survival probability is provided by the cut calibrated in the
data region with the most similar event topology, i.e. the DEP--based
cut for the \nubb~peak, the SEP--based cut and SP--based cut respectively for the Compton
continuum region at $2~\mbox{MeV}$ of $^{68}\mbox{Ga}$ and $^{60}\mbox{Co}$ (highlighted values). 
In cases when the MSE content of the region used for calibration is higher than in the studied events,
the calculated survival probability will be underestimated, and vice versa.
\begin{table}
   \caption{Percentage of survived events of the simulated \nubb--decays and internal
   sources of background after the cut in an $80~\mbox{keV}$ wide region
   centered on the $^{76}\mbox{Ge}$ \Qbb~of $2039~\mbox{keV}$. 
   The highlighted numbers represent the most realistic 
   predictions for experimental survival probabilities assuming experimental
   PSD cut with DEP fixed to 90\%.
   The uncertainties 
   given in parenthesis for the least significant digits include
   statistical as well as systematic uncertainties estimated by cut
   parameter variation. 
   The $^{60}\mbox{Co}$ result has an additional systematic uncertainty due 
   to a small difference on the event topology between the SP region and the 
   studied region.
   }
   \begin{center}
   \renewcommand{\arraystretch}{1.2}
   \setlength{\tabcolsep}{10pt}
   \begin{tabular}
      {| c |   
       c l @{ (}  c @{)} c 
       c l @{ (}  c @{)} c 
       c l @{ (}  c @{)} c |}
      \hline
      \multicolumn{1}{|c|}{source} 
      &\multicolumn{4}{c}{DEP fixed to 90\%} 
      &\multicolumn{4}{c}{SEP fixed to 5.45\%} 
      &\multicolumn{4}{c|}{SP  fixed to 0.08\%}\\
      \hline
      \nubb            && {\bf 0.86} & {\bf 3} &&&       0.77  &      3  &&&      0.57   & 13     &\\
      $^{68}\mbox{Ga}$ &&      0.067 &     11  &&& {\bf 0.045} & {\bf 3} &&&      0.032  & 7      &\\
      $^{60}\mbox{Co}$ &&      0.019 &      4  &&&     0.0130  &     11  &&& {\bf 0.009} & {$^{+4}_{-2}$}&\\
      \hline
   \end{tabular}
   \end{center}
   \label{tab:psdInt}
\end{table}

The \nubb~spectrum shows a peak at the \Qbb~energy and a tail extending to low
energies due to the events in which part of the total energy is either lost in the
dead layer or escape the detector.
Since the electrons generated by \nubb~decay events have a significant probability to emit a bremsstrahlung photon, 
not all the events in the \Qbb~peak and in the tail are SSE. 
The MSE contamination is similar to that in the DEP of $^{208}\mbox{Tl}$. 
However, as the energy of the \nubb~decay event is higher than in DEP 
events, also the probability of bremsstrahlung and in turn of an MSE interaction
is higher and the PSD cut survival probability is thus slightly lower.

The decay chain of the cosmogenic $^{68}\mbox{Ge}$ consists of an EC decay into
$^{68}\mbox{Ga}$, which then decays with $1.13~\mbox{h}$ half--life into
$^{68}\mbox{Zn}$ via EC ($11\%$) and $\beta^+$~decay ($89\%$).
The positron has an energy endpoint of $1.9~\mbox{MeV}$, thus to produce events
at the $^{76}\mbox{Ge}$ \Qbb ($2039~\mbox{keV}$) its absorption has to be
accompanied by an energy deposition from its annihilation photons.
Such events have a topology resembling that of SEP events (strong
electron/positron
interaction combined with a weaker $511~\mbox{keV}$ gamma--ray interaction)
and thus a comparable survival probability.

The $^{60}\mbox{Co}$ spectrum shows the two characteristic peaks at $1173.2$ and
$1332.5~\mbox{keV}$ generated in cascade. As the decay occurs inside the
detector, the probability of  a coincident detection is high, resulting in the
summation peak at $2505.7~\mbox{keV}$.  Moreover, the electron associated to the
$\beta^-$ decay of $^{60}\mbox{Co}$ to $^{60}\mbox{Ni}$
(endpoint energy of $318~\mbox{keV}$) is also absorbed leading to the peculiar
``triangular'' tail at the high-energy side of the peaks.  The events in the
\Qbb~region are a result of the cascade gamma and beta summation and
consequently their topologies are highly multi--site, leading to suppression by
a factor of 100.

\section{Conclusions and outlook}
In this paper we discussed the development of an integrated simulation
tool for the description of high-purity Ge detectors and its
application to derive the survival probabilities of neutrinoless
double decays of $^{76}\mbox{Ge}$ and of internal decays of
$^{68}\mbox{Ga}$ and $^{60}\mbox{Co}$ in BEGe type detectors.

The interaction points of the gamma--rays in the crystal are simulated
with the \GF~based MaGe~framework; the pulse shapes associated to the
interaction points are determined based on the electric field
calculation with an enhanced version of the MGS program. The
outputs of the two computations are combined together to determine the final pulse shape
corresponding to the interaction of the gamma rays within the
detector. This pulse shape is then convoluted with the front--end
electronics response and experimentally measured noise to produce
signals that can be directly compared to the experimental pulse
shapes.

The particular case of the BEGe detector chosen for the specific
application of the simulation is related to the future use of this
type of detectors in the Phase II of the \Gerda~neutrinoless double
beta decay experiment. Previous measurements performed with 
gamma--ray sources have evidenced an enhanced capability of these
detectors to discriminate between single--site and multiple--site
events from the analysis of the signal pulse shape.
The characteristic spread in time of the signal current pulse created
by the different charge clusters of a multi-site event has its origin
in the sharply rising weighting potential close to the small read--out
electrode.

The validation of the simulation was performed in two different
ways. In one of them the calculated pulses and their rise times are
compared to the measured ones generated in well defined positions by
the low--energy gamma rays of an $^{241}\mbox{Am}$ source. This method
is limited to the region near the outer surface of the Ge
crystal and strongly affected by the definition of the dead layer and
the size of the spatial grid in the calculation. The other is an
integral method and involves the comparisons of the rise time and $A/E$
distributions obtained from simulated and experimentally recorded interactions of high--energy
gamma--rays from a $^{208}\mbox{Tl}$ source with the
crystal. Such gamma rays interact in the whole volume of the detector
and the result of the comparison is a measure of the overall goodness
of the simulation. Both methods show a good agreement between the
measured and simulated data.  The slight deviations are mainly due to
inaccuracy in the geometry definition (especially important in case of
the p+ electrode), the finite grid width used for defining the library
of pulses and the charge carrier mobility model parametrization
(measured by using a different crystal at a different temperature).

The simulation was then used to estimate the capability of the pulse
shape discrimination method to reduce the gamma-ray background
produced by internal $^{68}~\mbox{Ga}$ and $^{60}\mbox{Co}$ sources in the energy 
region of the \nubb~decay. 
The $^{76}\mbox{Ge}$
\nubb~decay was also simulated in order to determine the PSD
acceptance of its signal. The results are presently dominated by
systematic errors related to analysis adjustments developed to
overcome the imperfections of the simulation, as the $A/E$ parameter is
particularly sensitive to such effects.  Nevertheless, the outcome
represents a reasonable prediction of the PSD background
discrimination efficiency for the \Gerda~experiment. The \nubb~decay
acceptance was found to be ($86\pm3$)\%, while the internal
backgrounds were drastically reduced. Only ($0.9^{+0.4}_{-0.2}$)\% of
$^{60}\mbox{Co}$ events and ($4.5\pm0.3$)\% of $^{68}\mbox{Ga}$ events
at \Qbb remain after the analysis cut. The $^{68}\mbox{Ga}$
suppression can be combined with the time coincidence rejection method
proposed in \cite{bib:gerdaProposal} which takes advantage of the short (1
hour) half life of $^{68}\mbox{Ga}$ and the $10~\mbox{keV}$ X--ray or
Auger electron emission ($\sim86\%$ probability) from its mother
isotope $^{68}\mbox{Ge}$. This latter method, providing reduction by
about a factor of 5, will also benefit from the low--noise
characteristic of BEGe detectors, maintaining a low--energy threshold
well below the interesting $10~\mbox{keV}$ X--ray energy. Owing to
the unique qualities of BEGe detectors, two of the most challenging
\Gerda~backgrounds, originating from internal $^{60}\mbox{Co}$ and
$^{68}\mbox{Ga}$ produced by cosmic ray interactions in germanium, 
can be reduced by a factor of 100 by analysis of the germanium signal only.

\acknowledgments 
The authors would like to thank Dino Bazzacco, Bernhard Schwingenheuer, Stefano
Riboldi and Enrico Borsato for the stimulating discussions and useful comments
on this work. 
They also acknowledge Matthias Laubenstein for providing help and supporting the
measurements in the low background laboratory.

This work was supported in part by the Transregio Sonderforschungsbereich
SFB/TR27 ``Neutrinos and Beyond'' by the Deutsche Forschungsgemeinschaft and by
the Munich Cluster of Excellence ``Origin and Structure of the Universe''.

%
%
%
%

\end{document}